\documentclass[12pt]{article}
\pdfoutput=1
\usepackage[nosort]{cite}

\usepackage{xcolor}

\usepackage{epsfig}
\usepackage{amsfonts}
\usepackage{amscd}
\usepackage{amsthm}
\usepackage{latexsym}
\usepackage{amsmath,amssymb}
\usepackage{verbatim}
\usepackage{setspace}
\usepackage{color}
\usepackage{fancyhdr}
\usepackage{cite}
\usepackage{hyperref}
\usepackage{tikz}
\usepackage{tikz-cd}
\usepackage{slashed}
\usepackage{soul}
\usepackage{multirow}
\usepackage{makecell}
\usetikzlibrary{calc}
\usetikzlibrary{topaths}
\usetikzlibrary{decorations}
\usetikzlibrary{decorations.pathmorphing}

\usepackage{draft}
\usepackage{hyperref}
\usepackage{graphicx,color,subfig}
\usepackage{cite}
\usepackage{mciteplus}
\usepackage{skak}

\numberwithin{equation}{section}

\newcommand{\lcm}{\operatorname{lcm}}

\newcommand{\Jac}{\operatorname{Jac}}

\def\({\left(}
\def\){\right)}

\begin{document}

\begin{titlepage}
	
\begin{center}

\hfill \\
\hfill MIT-CTP/5462, YITP-SB-2022-31\\
\vskip 1cm

\title{2+1d Compact Lifshitz Theory, Tensor Gauge Theory, and Fractons}

\author{Pranay Gorantla$^{1}$, Ho Tat Lam$^{2}$, Nathan Seiberg$^{3}$, and Shu-Heng Shao$^{4}$}

\address{${}^{1}$Physics Department, Princeton University}
\address{${}^{2}$Center for Theoretical Physics, Massachusetts Institute of Technology}
\address{${}^{3}$School of Natural Sciences, Institute for Advanced Study}
\address{${}^{4}$C.\ N.\ Yang Institute for Theoretical Physics, Stony Brook University}

\end{center}

\vspace{2.0cm}

\begin{abstract}\noindent
The 2+1d continuum Lifshitz theory of a free compact scalar field plays a prominent role in a variety of quantum systems in condensed matter physics and high energy physics. It is known that in compact space, it has an infinite ground state degeneracy.  In order to understand this theory better, we consider two candidate lattice regularizations of it using the modified Villain formalism.  We show that these two lattice theories have significantly different global symmetries (including a dipole global symmetry), anomalies, ground state degeneracies, and dualities. In particular, one of them is self-dual.  Given these theories and their global symmetries, we can couple them to corresponding gauge theories.  These are two different $U(1)$ tensor gauge theories.  The resulting models have excitations with restricted mobility, i.e., fractons.  Finally, we give an exact lattice realization of the fracton/lineon-elasticity dualities for the  Lifshitz theory, scalar and vector charge gauge theories.
\end{abstract}

\vfill
	
\end{titlepage}

\eject

\tableofcontents

\section{Introduction}\label{sec:intro}

In recent years, a large class of models with strange features have been discovered with important applications to condensed matter physics and quantum information. The most striking of these are the fracton models \cite{Chamon:2004lew,Haah:2011drr,Vijay:2016phm} which host particle-like excitations with restricted mobility, such as those that cannot move (fractons), or can move only in a line (lineons) or a plane (planeons). (See \cite{Nandkishore:2018sel,Pretko:2020cko,Grosvenor:2021hkn,Brauner:2022rvf,McGreevy:2022oyu,Cordova:2022ruw} for reviews on this subject.) These models defy a standard continuum quantum field theory description at low energies. Instead, their peculiar features are captured by non-standard field theories with exotic global symmetries. One important consequence of such symmetries is a large ground state degeneracy \cite{Haah:2020ghp}, which is infinite in the continuum limit.

The challenge to find a standard continuum low-energy description of these theories points to a missing deep insight in our understanding of continuum quantum field theory.  As we will demonstrate below, innocent-looking continuum models can be quite subtle and need a careful definition.  And their physical consequences depend sensitively on that definition.  In the opposite direction, some innocent-looking lattice models might not have any continuum low-energy field theory description.  It is expected that a detailed study of lattice models, continuum models, and the relations between them will enhance our understanding of these important issues.

Perhaps the simplest theory that exhibits this behavior is the 2+1d compact Lifshitz field theory \cite{Henley1997,Moessner2001,Vishwanath:2004,Fradkin:2004,Ardonne:2003wa,Ghaemi2005,Chen:2009ka,2018PhRvB..98l5105M,Yuan:2019geh,
Lake:2022ico} described by the Lagrangian\footnote{An even simpler version of this theory in 1+1d is analyzed in \cite{Gorantla:2022eem}.}
\ie\label{intro:lif-lag}
\mathcal L = \frac{\mu_0}{2} (\partial_\tau \phi)^2 + \frac{1}{2\mu} (\nabla^2 \phi)^2~,
\fe
where $\phi \sim \phi + 2\pi$ is a compact scalar, and $\nabla^2 = \partial_x^2 + \partial_y^2$ is the 2d spatial Laplacian operator. (Throughout this paper, we work in the Euclidean signature, and $\tau$ denotes Euclidean time.) The compact Lifshitz field theory has appeared in many different physical contexts, including deconfined quantum criticality \cite{Vishwanath:2004}.

It is well-known that, on a 2d spatial torus with periodic boundary conditions, this theory has infinite ground state degeneracy, where the ground states are labelled by the winding numbers in the $x$ and $y$ directions \cite{Henley1997,Moessner2001}. This is a consequence of the dipole symmetry \cite{Griffin:2013dfa,Griffin:2014bta,Pretko:2016kxt,Pretko:2016lgv,Pretko:2018jbi,Gromov:2018nbv,Seiberg:2019vrp,Shenoy:2019wng,
Gromov:2020rtl,Gromov:2020yoc,Chen:2020jew,Du:2021pbc,Stahl:2021sgi,Lake:2022ico,Gorantla:2022eem,Jensen:2022iww,Kapustin:2022fzp} which shifts $\phi$ by \emph{linear functions} in $x$ and $y$. On the other hand, on the plane $\mathbb R^2$, the symmetry is much larger and includes shifts of $\phi$ by \emph{harmonic functions} of $x$ and $y$, of which the linear functions form only a tiny subset. To regularize this infinity and make sense of the exotic symmetry, we wish to place the theory on the lattice.

It is commonly the case that given a continuum theory, there are several ways to regularize it on the lattice. The difference between them is in irrelevant operators in the continuum theory and therefore, it is not important.  As we will see, this is not the case here.

How should we regularize the 2+1d Lifshitz Lagrangian \eqref{intro:lif-lag}? One could ``discretize'' the Laplacian operator $\nabla^2$ as $\Delta_x^2 + \Delta_y^2$, which is the discrete Laplacian operator on the 2d spatial torus lattice.\footnote{We label the sites on the lattice as $(x,y)$, where $x,y$ are integers. (At the risk of confusing the reader, we use the same characters $(x,y)$ for the continuum coordinates and the discrete lattice coordinates.) We define $\Delta_x f(x+\frac{1}{2},y)\equiv f(x+1,y)-f(x,y)$ for a function on the sites, $\Delta_x f_x(x,y) \equiv  f_x(x+\frac{1}{2},y) - f_x(x-\frac{1}{2},y)$ for a function on the $x$-links, $\Delta_x f_y(x+\frac{1}{2},y+\frac{1}{2}) \equiv f_y(x+1,y+\frac{1}{2}) - f_y(x,y+\frac{1}{2})$ for a function on $y$-links, and $\Delta_x f_{xy}(x,y+\frac{1}{2}) \equiv  f_{xy}(x+\frac{1}{2},y+\frac{1}{2}) - f_{xy}(x-\frac{1}{2},y+\frac{1}{2})$ for a function on plaquettes. $\Delta_y$ is defined in a similar way. Using these definitions, if $f$ is a function on the sites, then $\Delta_x f$ is a function on the $x$-links, and so $\Delta_x^2 f(x,y)= f(x+1,y)-2f(x,y)+f(x-1,y)$ and $\Delta_x \Delta_y f(x+\tfrac12,y+\tfrac12) = f(x+1,y+1) - f(x,y+1) - f(x+1,y) + f(x,y)$. We follow similar notation for functions that further depend on $\tau$.} This discretization has the advantage of being well-defined on other spatial lattices, including general graphs. (See \cite{Gorantla:2022mrp} for a discussion of exotic lattice models on graphs based on the discrete Laplacian operator.)
Alternatively, one could first integrate by parts and replace $(\nabla^2\phi)^2$ with
\ie\label{2dphi-second-discretization}
(\partial_x^2 \phi)^2 + (\partial_y^2\phi)^2 + 2(\partial_x \partial_y \phi)^2~,
\fe
and then discretize the three terms separately.  We will see that these two discretizations are not the same and they do not defer merely by irrelevant operators.  Instead, the discretization following from \eqref{2dphi-second-discretization} is very close to the continuum theory \eqref{intro:lif-lag}.  But the other one is an interesting peculiar lattice model, whose relation to the continuum theory is unclear.

Since $\phi$ is compact, the discretization has to take that into account.  One possibility is to use trigonometric functions in the lattice action. We prefer to use a Villain-type formalism \cite{Villain:1974ir}.  Here, we introduce on the lattice integer-valued gauge fields.  Then, the two different discretizations above lead to the following schematic spatial kinetic terms:
\ie\label{intro:cosine}
\text{Laplacian $\phi$-theory:}&\quad
[(\Delta_x^2 + \Delta_y^2) \phi - 2\pi n]^2+\cdots~,
\\
\text{Dipole $\phi$-theory:}&\quad
(\Delta_x^2 \phi - 2\pi n_{xx})^2 + (\Delta_y^2 \phi- 2\pi n_{yy})^2 + 2(\Delta_x \Delta_y \phi -2\pi n_{xy})^2 +\cdots~.
\fe
(We defer a detailed discussion of the Villain integer gauge fields and the additional terms in the ellipses to Section \ref{sec:2dphi}.)

We refer to the first model as the Laplacian $\phi$-theory because it makes use of the discrete Laplacian operator, and the second model as the dipole $\phi$-theory because it has dipole global symmetries, which will be discussed in Section \ref{sec:2ddipphi}.

Following \cite{Seiberg:2019vrp,paper1,paper2,paper3,Gorantla:2020xap,Gorantla:2020jpy,Rudelius:2020kta,Gorantla:2021svj,Gorantla:2021bda,
Burnell:2021reh,Gorantla:2022eem,Gorantla:2022mrp}, we focus on the global symmetries and other global aspects. To this end, we analyze the modified Villain formulation \cite{Sulejmanpasic:2019ytl,Gorantla:2021svj} of these two lattice models. The advantage of this formulation is that the symmetries, anomalies, and dualities of the continuum theory are already manifest on the lattice.

While naively the two lattice models in \eqref{intro:cosine} are regularizations of the same continuum Lifshitz Lagrangian \eqref{intro:lif-lag}, they have very different properties:
\begin{itemize}
\item Their global symmetries are different. For example, the Laplacian model is invariant under the schematic transformation $\phi\to \phi +c xy$ with a constant $c$, while the dipole model does not enjoy such a symmetry.\footnote{The global aspects of such global symmetries will be discussed in details in Section \ref{sec:2dphi}. See, in particular, the discussion around equation \eqref{xyshift}.}
\item The ground state degeneracies (GSD) of the two models, when placed on a 2d spatial torus lattice with $L_i$ sites in the $i=x,y$ direction, are drastically different\footnote{Observe that the GSD of the Laplacian $\phi$-theory grows exponentially in the number of sites $L_x L_y$. A similar behavior is already present in a trivial system with decoupled spins. As explained in \cite{Gorantla:2022mrp} and reviewed in Section \ref{sec:2dlapphi-sym}, the origins of these two exponential behaviors are not the same.  In the Laplacian $\phi$-theory, it is the large orders of some of the symmetry generators, whereas in the decoupled spin system, it is the extensive number of symmetry generators.}
\ie
\text{Laplacian $\phi$-theory:}&\quad \text{GSD} \sim \exp\left( {4G\over \pi} L_x L_y\right)~,\quad L_i\to\infty ~,
\\
\text{Dipole $\phi$-theory:}&\quad \text{GSD}=L_xL_y~,
\fe
where $G\sim 0.916$ is the Catalan constant, for the Laplacian model.
\item The Laplacian model is not robust against perturbations of (momentum) symmetry-preserving local operators, while the dipole model is robust.
\item They have different dualities. The Laplacian lattice model is self-dual,\footnote{The authors of \cite{Vishwanath:2004} argued that the continuum Lifshitz theory is self-dual.  The corresponding lattice Laplacian theory has such a self-duality \cite{Gorantla:2022mrp}, which will be discussed in Section \ref{sec:selfdual}} but the dipole lattice model is dual to the lattice version of the vector charge theory \cite{Xu:2006,Pretko:2016kxt,Pretko:2016lgv,Bulmash:2018lid,Pretko:2018jbi,Oh:2021gee,Oh:2022klh}.  In the continuum, this tensor gauge theory has gauge fields $(\hat A_{\tau i} ,\hat A_{ij})$ and a spatial vector gauge parameter $\hat \alpha_i$ with gauge transformations
\ie
&\text{Vector charge theory:}\\
&\qquad\hat A_{\tau i} \sim \hat A_{\tau i} + \partial_\tau \hat \alpha_i~,\quad \hat A_{xy} \sim \hat A_{xy} + \partial_x \hat \alpha_y + \partial_y \hat \alpha_x~,\quad \hat A_{ii} \sim \hat A_{ii} + \partial_i \hat \alpha_i~,~~i=x,y\, .
\fe
This theory has defects that represent the worldline of particles that are allowed to move in one direction in space, i.e., it is a theory of lineons.
The  duality between the dipole $\phi$-theory and the vector charge theory had been discussed in the continuum in the context of elasticity theory \cite{Manoj:2020bcz}. We refer to this duality as the lineon-elasticity duality.  We will discuss its lattice version in Section \ref{sec:lineonelasticity}.

\end{itemize}

Other differences between the two models are  discussed in Section \ref{sec:2dphi}. The fact these two different regularizations of the same continuum Lifshitz Lagrangian \eqref{intro:lif-lag} have completely different global symmetries, GSD, and dualities highlights the ambiguity of working just with the naive continuum Lagrangian.  Having said that, it is clear that the dipole lattice theory is closer to the continuum Lifshitz theory than the more exotic Laplacian lattice theory.

The pure $U(1)$ gauge theories associated with the global symmetries of the two models of \eqref{intro:cosine} are also strikingly different. Schematically, the one associated with the Laplacian model has only two gauge fields, $A_\tau$ and $A$, with Laplacian gauge transformations
\ie
&\text{Laplacian gauge theory:}\\
&\qquad A_\tau \sim A_\tau + \partial_\tau \alpha~,\qquad A \sim A + \nabla^2 \alpha~.
\fe
On the other hand, the one associated with the dipole model is a rank-2 tensor gauge theory which has been discussed extensively in the literature \cite{Pretko:2016kxt,Pretko:2016lgv,Pretko:2018qru,Bulmash:2018lid,Slagle:2018kqf,Pretko:2018jbi,Pretko:2019omh,
Shenoy:2019wng,Oh:2021gee,Oh:2022klh}. It has four gauge fields, $A_\tau$ and $A_{xx},A_{yy},A_{xy}$, with rank-2 tensor gauge transformations
\ie
&\text{Scalar charge theory:}\\
&\qquad A_\tau \sim A_\tau + \partial_\tau \alpha~,\qquad A_{ij} \sim A_{ij} + \partial_i \partial_j \alpha~,\quad i,j=x,y~.
\fe
The gauge theory of $(A_\tau,A_{ij})$ (possibly coupled to matter fields) is referred to  in \cite{Pretko:2016kxt,Pretko:2016lgv,Bulmash:2018lid,Pretko:2018jbi} as the scalar charge theory to emphasize the fact that the gauge parameter $\alpha$ is a scalar. Similar to the Lifshitz theory, the continuum gauge transformations and Lagrangians do not  specify unambiguously all the global aspects (such as GSD discussed below) of these gauge theories. In Section \ref{sec:2dA}, we regularize these two gauge theories using the (modified) Villain lattice formulation, while preserving all the global symmetries, dualities, and anomalies. From this point on, we refer to the (modified) Villain lattice versions of these two gauge theories as the Laplacian gauge theory and the scalar charge theory.

Both the 2+1d Laplacian and the scalar charge theories are natural generalizations of the 1+1d tensor gauge theory studied in \cite{Gorantla:2022eem}. Both gauge theories have line defects that describe the worldline of an immobile particle, i.e., they are fracton models.

However, these models do not have GSD that grows sub-extensively in the system size. Their GSDs on a 2d spatial torus lattice are:\footnote{To be precise, one can write a $\theta$-term in the Laplacian gauge theory, where $\theta \sim \theta + 2\pi$. When $\theta \ne \pi$, the ground state is non-degenerate, whereas there is a two-fold degeneracy at $\theta = \pi$.}
\ie
\text{Laplacian gauge theory:}&\quad\text{GSD}=1~,\\
\text{Scalar charge theory:}&\quad\text{GSD}=\text{gcd}(L_x,L_y)~.
\fe
Interestingly, the GSD of the scalar charge theory of $(A_\tau,A_{ij})$ depends on $\text{gcd}(L_x,L_y)$, which is a manifestation of UV/IR mixing in these exotic models \cite{paper1,Gorantla:2021bda}.

\begin{figure}[t]
\begin{center}
\begin{tikzcd}[row sep=huge,column sep=large]
\phi_\text{dip} \arrow[dash]{d}[swap]{\substack{\text{lineon-elasticity}\\\text{duality}}} \arrow{dr}[swap]{} \arrow[white]{r}[swap,black,inner sep=0.4cm]{\text{Higgs}} & \hat \phi_i \arrow[dash]{d}[]{\substack{\text{fracton-elasticity}\\\text{duality}}}\arrow{dl}[]{}
\\
(\hat A_{\tau i},\hat A_{ij}) & (A_\tau,A_{ij})
\end{tikzcd}
$\qquad \qquad$
\begin{tikzcd}[baseline=0.7cm,row sep=normal,column sep=large]
\phi_\text{Lap} \arrow[dash,loop,"\text{self-duality}"'] \arrow{d}[]{\text{Higgs}}
\\
(A_\tau,A)
\end{tikzcd}
\end{center}
\caption{\textbf{Left:} Dualities and relations between the two matter theories, the dipole $\phi$-theory and the $\hat\phi_i$-theory, and the two gauge theories, the scalar charge theory $(A_\tau,A_{ij})$ and the vector charge theory $(\hat A_{\tau i},\hat A_{ij})$. The dipole $\phi$-theory is the Higgs field of the $(A_\tau,A_{ij})$ theory, or conversely, the $(A_\tau,A_{ij})$ theory gauges the momentum dipole symmetry of $\phi$. The relation between $\hat\phi_i$ and $(\hat A_{\tau i}, \hat A_{ij})$ is similar. The duality in Section \ref{sec:fractonelasticity} between $\hat \phi_i$ and $(A_\tau, A_{ij})$ is the lattice version of the fracton-elasticity duality of \cite{Pretko:2018qru,Pretko:2019omh,Nguyen:2020yve}, while the duality in Section \ref{sec:lineonelasticity} between the dipole $\phi$-theory and $(\hat A_{\tau i}, \hat A_{ij})$ is the lattice version of the lineon-elasticity duality of \cite{Manoj:2020bcz}. \textbf{Right:} The Laplacian $\phi$-theory is self-dual \cite{Vishwanath:2004,Gorantla:2022mrp} (see Section \ref{sec:selfdual}). The Laplacian $\phi$-theory Higgses the Laplacian gauge theory $(A_\tau, A)$, which has no duality. Note that while the Laplacian $\phi$-theory (denoted as $\phi_\text{Lap}$) and the dipole $\phi$-theory (denoted as $\phi_\text{dip}$) are both naive discretizations of the same continuum Lifshitz Lagrangian \eqref{intro:lif-lag}, their dualities are drastically different. Even though we use the continuum notations for these fields, our dualities are established as exact lattice dualities. }\label{fig:duality}
\end{figure}
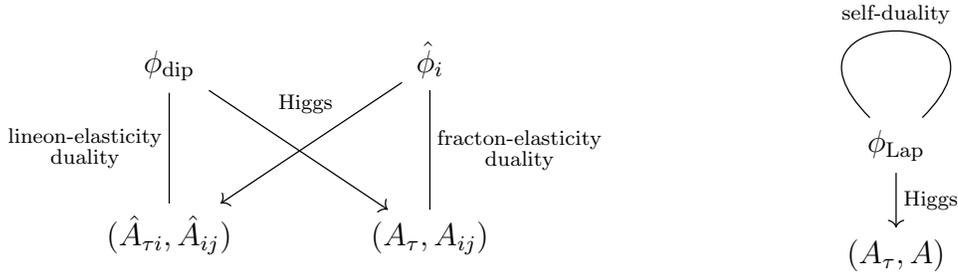

The two models also differ in other ways. The scalar charge theory is exactly dual to the elasticity theory with displacement fields $(\hat \phi_x, \hat\phi_y)$. The continuum version of this exact lattice duality between the scalar charge theory and the $\hat\phi_i$-theory, known as the fracton-elasticity duality, has been discussed in \cite{Pretko:2018qru,Pretko:2019omh,Nguyen:2020yve}. See Figure \ref{fig:duality} for some of the dualities and relations between these models. In contrast, the Laplacian gauge theory does not enjoy any duality. Other differences between the two theories are   discussed in Section \ref{sec:2dA}.

The Laplacian $\phi$-theory and gauge theory were recently introduced in \cite{Gorantla:2022mrp} on a general spatial graph. In this paper, we will place them on a 2d torus, and compare them with their dipole counterparts: the dipole $\phi$-theory and the scalar charge theory. We will make use of the results from \cite{Gorantla:2022mrp} throughout.

The various 2+1d theories analyzed in this paper, and their schematic continuum Lagrangians, are summarized in Table \ref{tbl:contlag}.

\renewcommand{\arraystretch}{1.5}
\begin{table}[h!]
\begin{center}
\begin{tabular}{|c|c|}
\hline
Theory & Continuum Lagrangian
\tabularnewline
\hline
\noalign{\vskip 5pt}
\hline
Compact Lifshitz theory $\phi$ \cite{Henley1997,Moessner2001,Vishwanath:2004,Fradkin:2004,Ardonne:2003wa,Ghaemi2005,Chen:2009ka,2018PhRvB..98l5105M,Yuan:2019geh,
Lake:2022ico} & $\mathcal L = (\partial_\tau \phi)^2 + (\nabla^2 \phi)^2$
\tabularnewline
\hline
\noalign{\vskip 5pt}
\hline
Elasticity theory of $\hat \phi_i$ \cite{Pretko:2018qru,Pretko:2019omh,Nguyen:2020yve,Chen:2020jew} & $\mathcal L = \sum_i (\partial_\tau \hat \phi_i)^2 + \sum_{i,j} (\partial_i \hat \phi_j + \partial_j \hat \phi_i)^2$
\tabularnewline
\hline
\noalign{\vskip 5pt}
\hline
Scalar charge theory \cite{Pretko:2016kxt,Pretko:2016lgv,Bulmash:2018lid,Pretko:2018jbi} & $\mathcal L = \sum_{i,j} E_{ij}^2 + \sum_i B_i^2$
\tabularnewline
\cline{1-2}
$A_\tau\sim A_\tau + \partial_\tau \alpha$ & $E_{ij} = \partial_\tau A_{ij} - \partial_i \partial_j A_\tau$
\tabularnewline
$A_{ij}\sim A_{ij} + \partial_i \partial_j \alpha$ & $B_i = \partial_i A_{jj} - \partial_j A_{ij}~,~i\ne j$
\tabularnewline
\hline
\noalign{\vskip 5pt}
\hline
Vector charge theory \cite{Pretko:2016kxt,Pretko:2016lgv,Bulmash:2018lid,Pretko:2018jbi} & $\mathcal L = \sum_{i,j} \hat E_{ij}^2 + \hat B^2$
\tabularnewline
\cline{1-2}
$\hat A_{\tau i} \sim \hat A_{\tau i} + \partial_\tau \hat \alpha_i$ & $\hat E_{ii} = \partial_\tau \hat A_{ii} - \partial_i \hat A_{\tau i}$
\tabularnewline
$\hat A_{ii} \sim \hat A_{ii} + \partial_i \hat \alpha_i$ & $\hat E_{xy} = \partial_\tau \hat A_{xy} - \partial_x \hat A_{\tau y} - \partial_y \hat A_{\tau x}$
\tabularnewline
$\hat A_{xy} \sim \hat A_{xy} + \partial_x \hat \alpha_y + \partial_y \hat \alpha_x$ & $\hat B = \partial_x^2 \hat A_{yy} + \partial_y^2 \hat A_{xx} - \partial_x \partial_y \hat A_{xy}$
\tabularnewline
\hline
\noalign{\vskip 5pt}
\hline
Laplacian gauge theory \cite{Gorantla:2022mrp} & $\mathcal L = E^2$
\tabularnewline
\cline{1-2}
$A_\tau \sim A_\tau + \partial_\tau \alpha$ & \multirow{2}{*}{$E = \partial_\tau A - \nabla^2 A_\tau$}
\tabularnewline
$A \sim A + \nabla^2 \alpha$ &
\tabularnewline
\hline
\end{tabular}
\end{center}
\caption{Summary of various 2+1d theories analyzed in this paper, and their naive continuum Lagrangians. Here, $i,j=x,y$, and repeated indices are not summed over.}\label{tbl:contlag}
\end{table}

The rest of this paper is organized as follows. In Section \ref{sec:2dphi}, we study the modified Villain versions of the Laplacian and dipole models of \eqref{intro:cosine}. One of the main results here is the stark contrast between the scalings of their GSDs with the system size.  While the GSD of the dipole model grows linearly in the system size, the GSD of the Laplacian model grows exponentially. This can be understood as a consequence of the distinct exotic global symmetries of the two models. These differences are summarized in Table \ref{tbl:2dphi}. Interestingly, the modified Villain version of the dipole model realizes the lineon-elasticity duality exactly on the lattice.

In Section \ref{sec:2dA}, we discuss the modified Villain versions of the Laplacian gauge theory and the scalar charge (gauge) theory. Once again, these two theories differ in several global aspects. For example, while both theories have fracton defects, a dipole of fractons is mobile in the scalar charge theory, but it is immobile in the Laplacian gauge theory. This can also be understood as a consequence of the distinct exotic global (time-like) symmetries of the two theories. These differences are summarized in Table \ref{tbl:2dA}. It is interesting to note that the modified Villain version of the scalar charge theory realizes the fracton-elasticity duality exactly on the lattice.

In Appendix \ref{app:2ddip}, we give more details about several aspects of the dipole $\phi$-theory and the scalar charge theory. These include the global symmetries, symmetry operators, charged operators, and defects in both theories.

In Appendix \ref{app:poly}, we explore further properties of the the Laplacian theories on the square lattice. Specifically, we discuss the naturalness of the Laplacian $\phi$-theory, and the immobility of any finite set of defects in the Laplacian gauge theory on the square lattice.

\section{2+1d compact Lifshitz theory}\label{sec:2dphi}

We begin with a naive analysis of the global symmetries of the continuum 2+1d compact Lifshitz theory \cite{Henley1997,Moessner2001,Vishwanath:2004,Fradkin:2004,Ardonne:2003wa,Ghaemi2005,Chen:2009ka,2018PhRvB..98l5105M,Yuan:2019geh,
Lake:2022ico} on a Euclidean spacetime $3$-torus with lengths $\ell_\tau$, $\ell_x$, and $\ell_y$. As we will see, the dipole global symmetry has infinite order, which leads to infinite ground state degeneracy. This calls for a lattice regularization.

The Lagrangian is
\ie\label{2ddipphi-cont-lag}
\mathcal L = \frac{\mu_0}{2} (\partial_\tau \phi)^2 + \frac{1}{2\mu} (\nabla^2 \phi)^2~,
\fe
where $\phi$ is a compact scalar, i.e., $\phi \sim \phi + 2\pi$, and $\mu_0$ and $\mu$ are coupling constants of mass dimension $1$.

The global symmetry of this theory includes:
\begin{itemize}
\item a $U(1)$ momentum (shift) symmetry, $\phi \rightarrow \phi + c$, where $c \sim c+2\pi$ is a circle-valued constant,\footnote{More precisely, here and throughout, by saying that $c$ is circle-valued we mean $c\in \mathbb{R}/2\pi \mathbb{Z}$. }
\item a $\mathbb Z \times \mathbb Z$ momentum dipole symmetry,
\ie
\phi \rightarrow \phi + \frac{2\pi m_x x}{\ell_x} + \frac{2\pi m_y y}{\ell_y}~,
\fe
where $m_x,m_y\in\mathbb Z$, and $x,y$ are continuum coordinates (rather than integers labelling the sites on a lattice), and
\item a $U(1)$ one-form symmetry \cite{Gaiotto:2014kfa}, which will be referred to as the ``winding dipole'' symmetry, with Noether currents\footnote{We refer to this ordinary one-form global symmetry as a ``winding dipole'' symmetry because it is the continuum limit of the $\mathbb{Z}_{L_x}\times \mathbb{Z}_{L_y}$ winding dipole symmetry \eqref{dipolephi:windingdipole} of the dipole $\phi$-theory on a torus. See \cite{Gorantla:2022eem} for a similar phenomenon in 1+1d.}
\ie
J_{[\tau x]} = \frac{1}{2\pi} \partial_y \phi~,\qquad J_{[\tau y]} = -\frac{1}{2\pi} \partial_x \phi~,\qquad J_{[xy]} = \frac{1}{2\pi} \partial_\tau \phi~,
\fe
that obey the current conservation equation
\ie
\partial^\mu J_{[\mu\nu]}=0~.
\fe
The conserved current leads to two conserved, integer-valued charges:
\ie\label{2ddipphi-U1wind1form}
Q_x = \oint dy~J_{[ \tau x]}~,\qquad Q_y=\oint dx~J_{[\tau y]}~.
\fe
Note that this symmetry is not a subsystem symmetry because $Q_x$ is independent of $x$ and $Q_y$ is independent of $y$.    In addition, as with all one-form global symmetries, it leads to a time-like symmetry.\footnote{Time-like global symmetries are symmetries that do not act on the ordinary Hilbert space, but do act on the Hilbert space of defects.  They provide a convenient and powerful way to relate the restricted mobility of defects to a global symmetry.  Consequently, they control the restricted mobility of charged excitations.  See \cite{Gorantla:2022eem} for more details.}
\end{itemize}
This theory is also invariant under translations, spatial rotations, and the Lifshitz scale transformation $\tau \rightarrow \lambda^2 \tau$, $x\rightarrow \lambda x$, and $y\rightarrow \lambda y$.

The $\mathbb Z\times \mathbb Z$ momentum dipole symmetry operators do not commute with the winding dipole charges because the momentum dipole symmetry shifts the field $\phi$ by a configuration charged under the winding dipole symmetry.\footnote{This lack of commutativity between the momentum dipole symmetry and the winding dipole symmetry means that the full symmetry group is realized projectively.  This fact can be thought of as a mixed anomaly between these two symmetries.} This leads to an infinite ground state degeneracy. To regularize this infinity, we wish to place the theory on the lattice and study its modified Villain version \cite{Sulejmanpasic:2019ytl,Gorantla:2021svj}.

As mentioned in the introduction, there are two natural ways to ``discretize'' the continuum Lagrangian \eqref{2ddipphi-cont-lag}. In the next two subsections, we study the modified Villain model associated with these two discretizations in detail, and note important distinctions between them. These differences are summarized in Table \ref{tbl:2dphi}.

\renewcommand{\arraystretch}{2}
\begin{table}[h!]
\begin{center}
\begin{tabular}{|c|c|c|c|}
\hline
Theory & Laplacian $\phi$-theory & Dipole $\phi$-theory & \makecell{Continuum compact\\Lifshitz theory}
\tabularnewline
\hline
Duality & Self-dual & \makecell{Dual to the vector\\charge theory: $(\hat {\cal A}_{\tau i} ,\hat {\cal A}_{ij})$} & ?
\tabularnewline
\hline
\multirow{2}{*}{\makecell{Space-like\\global symmetry}} & $U(1) \times \Jac$ mom. & $U(1)\times \mathbb Z_{L_x} \times \mathbb Z_{L_y}$ mom. & $U(1) \times \mathbb Z^2$ mom.
\tabularnewline
\cline{4-4}
& $U(1) \times \Jac$ wind. & $U(1)^3 \times \mathbb Z_{L_x} \times \mathbb Z_{L_y}$ wind. & \multirow{2}{*}{\makecell{$U(1)$ one-form \\wind. dipole}}
\tabularnewline
\cline{1-3}
\makecell{Time-like\\global symmetry} & does not exist & $U(1)^2 \times \mathbb Z_{\gcd(L_x,L_y)}$ &
\tabularnewline
\hline
\makecell{Ground state\\degeneracy} & $|\Jac|\sim \exp\left(\frac{4G}{\pi} L_x L_y\right)$ & $L_x L_y$ & $\infty$
\tabularnewline
\hline
Robustness & No & Yes & ?
\tabularnewline
\hline
\end{tabular}
\end{center}
\caption{Comparison of the two ``discretizations'' of the 2+1d continuum compact Lifshitz Lagrangian \eqref{2ddipphi-cont-lag}. Here, ``$\Jac$'' is short-hand for the Jacobian group $\Jac(C_{L_x} \times C_{L_y})$ of the 2d torus graph $C_{L_x} \times C_{L_y}$, ``mom.'' stands for momentum, ``wind.'' stands for winding, and $G$ is the Catalan constant. The last row refers to robustness of the winding symmetry and ground state degeneracy after imposing the momentum symmetry---if the momentum symmetry is not imposed, then both models are not robust.}\label{tbl:2dphi}
\end{table}

\subsection{Laplacian $\phi$-theory}\label{sec:2dlapphi}
In this subsection, we discuss the modified Villain model associated with the first discretization of $(\nabla^2\phi)^2$. We refer to it as the 2+1d Laplacian $\phi$-theory. (See \cite{Gorantla:2022mrp} for the discussion of this model on a general spatial graph.)

Let us place the theory on a periodic 3d Euclidean spacetime lattice with $L_\tau$, $L_x$, and $L_y$ sites. The modified Villain action of the 2+1d Laplacian $\phi$-theory is
\ie\label{2dlapphi-modVill-action}
S &= \frac{\beta_0}{2} \sum_{\tau\text{-link}} \left(\Delta_\tau \phi- 2\pi n_\tau \right)^2 + \frac{\beta}{2} \sum_\text{site} \left[ (\Delta_x^2+\Delta_y^2) \phi - 2\pi n \right]^2
\\
& \qquad + i\sum_{\tau\text{-link}} \tilde \phi \left[ \Delta_\tau n - (\Delta_x^2 + \Delta_y^2) n_\tau \right]~,
\fe
where $\phi$ is a real-valued field, $(n_\tau,n)$ are integer gauge fields, and $\tilde \phi$ is a real-valued Lagrange multiplier that makes the integer gauge fields flat. The locations of these fields are summarized in Table \ref{tbl:2dphiloc}. There is a gauge symmetry
\ie\label{2dlapphi-gaugesym}
&\phi \sim \phi + 2\pi k~,\qquad&&n_\tau \sim n_\tau + \Delta_\tau k~,
\\
&\tilde \phi \sim \tilde \phi + 2\pi \tilde k~,\qquad&&n \sim n + (\Delta_x^2+\Delta_y^2) k~,
\fe
where $k$ and $\tilde k$ are integer gauge parameters. This integer gauge symmetry makes the scalar fields $\phi$ and $\tilde \phi$ compact.

\renewcommand{\arraystretch}{1.5}
\begin{table}
\begin{center}
\begin{tabular}{|c|c|c|}
\hline
Location & Laplacian $\phi$-theory & Dipole $\phi$-theory
\tabularnewline
\hline
site & $\phi, n, \tilde n_\tau$ & $\phi,n_{ii},\hat n_{ii}$
\tabularnewline
\hline
$\tau$-link & $\tilde \phi,n_\tau,\tilde n$ & $\hat{\mathcal A}_{ii},n_\tau,\hat n$
\tabularnewline
\hline
$i$-link & -- & $\hat{\mathcal A}_{\tau i}$
\tabularnewline
\hline
$\tau i$-plaq & -- & --
\tabularnewline
\hline
$xy$-plaq & -- & $n_{xy},\hat n_{\tau xy}$
\tabularnewline
\hline
cube & -- & $\hat{\mathcal A}_{xy}$
\tabularnewline
\hline
\end{tabular}
\end{center}
\caption{Locations of various fields of the Laplacian and dipole $\phi$-theories of Section \ref{sec:2dphi}. Here, $i=x,y$ and repeated indices are not summed over.}\label{tbl:2dphiloc}
\end{table}

\subsubsection{Self-duality}\label{sec:selfdual}
The 2+1d Laplacian $\phi$-theory \eqref{2dlapphi-modVill-action} is self-dual with $\phi \leftrightarrow \tilde \phi$ and $\beta_0 \leftrightarrow \frac{1}{(2\pi)^2\beta}$. Indeed, using the Poisson resummation formula for the integers $n_\tau,n$, the dual action is
\ie
S &= \frac{1}{2(2\pi)^2 \beta} \sum_\text{site} (\Delta_\tau \tilde \phi- 2\pi \tilde n_\tau )^2 + \frac{1}{2(2\pi)^2\beta_0} \sum_{\tau\text{-link}} [ (\Delta_x^2+\Delta_y^2) \tilde \phi - 2\pi \tilde n ]^2
\\
& \qquad - i\sum_\text{site} \phi \left[ \Delta_\tau \tilde n - (\Delta_x^2 + \Delta_y^2) \tilde n_\tau \right]~,
\fe
where $(\tilde n_\tau,\tilde n)$ are integer gauge fields that make $\tilde \phi$ compact. Under the gauge symmetry \eqref{2dlapphi-gaugesym}, they transform as
\ie
\tilde n_\tau \sim \tilde n_\tau + \Delta_\tau \tilde k~,\qquad \tilde n \sim \tilde n + (\Delta_x^2 + \Delta_y^2) \tilde k~.
\fe
This self-duality was suggested in the continuum in \cite{Vishwanath:2004} and formulated, more recently, as an exact duality on the lattice in \cite{Gorantla:2022mrp}.

\subsubsection{Global symmetry and ground state degeneracy}\label{sec:2dlapphi-sym}
Let us discuss the global symmetry of the 2+1d Laplacian $\phi$-theory \eqref{2dlapphi-modVill-action} \cite{Gorantla:2022mrp}.
\begin{itemize}
\item There is a $U(1)$ momentum symmetry that acts as $\phi \rightarrow \phi + c$, where $c\sim c+2\pi$ is a circle-valued constant. A typical charged operator is $e^{i\phi}$.

\item There is a discrete momentum symmetry that acts as
\ie\label{2dlapphi-mom-har}
&\phi \rightarrow \phi + f(x,y)~,
\\
&n \rightarrow n + \frac{1}{2\pi} (\Delta_x^2 + \Delta_y^2) f(x,y)~,
\fe
with $f(x,y)$ any function satisfying $(\Delta_x^2 + \Delta_y^2) f(x,y) \in 2\pi \mathbb Z$. (We do not refer to it as a dipole symmetry, because the shift function $f(x,y)$ can have a more general form than for a dipole symmetry, where it is linear.) In other words, $f(x,y)$ is  a circle-valued discrete harmonic function on the 2d spatial torus lattice. The symmetry group formed by such functions is the \emph{Jacobian group}, $\Jac(C_{L_x} \times C_{L_y})$, of the 2d torus lattice $C_{L_x}\times C_{L_y}$ with $L_i$ number of sites in the $i$ direction. See \cite{BSMF:97,BAKER2007} for the definition of this group and \cite{Gorantla:2022mrp} for its relation to the Laplacian $\phi$-theory. There is no simple closed form formula for $\Jac(C_{L_x} \times C_{L_y})$; for example, $\Jac(C_2 \times C_2)=\mathbb{Z}_2\times \mathbb{Z}_2\times \mathbb{Z}_8$ and $\Jac(C_3 \times C_3)=\mathbb{Z}_6\times \mathbb{Z}_6\times \mathbb{Z}_{18}\times \mathbb{Z}_{18}$.

\item There is a $U(1)$ winding symmetry that acts as $\tilde \phi \rightarrow \tilde \phi + \tilde c$, where $\tilde c\sim \tilde c+2\pi$ is a circle-valued constant. A typical charged operator is $e^{i\tilde \phi}$.

\item There is a $\Jac(C_{L_x} \times C_{L_y})$ discrete winding symmetry that acts as
\ie
\tilde \phi \rightarrow \tilde \phi + \tilde f(x,y)~,
\fe
where $\tilde f(x,y)$ is a circle-valued discrete harmonic function on the 2d spatial torus lattice.
\end{itemize}
The discrete momentum and winding symmetries do not commute with each other, which leads to a large ground state degeneracy equal to the order of the Jacobian group, $|\Jac(C_{L_x} \times C_{L_y})|$. As shown in \cite{Gorantla:2022mrp}, the logarithm of GSD grows as $L_x L_y$. More concretely,
\ie
\log \text{GSD} \approx \frac{4G}{\pi} L_x L_y~,
\fe
where $G$ is the Catalan constant.  More generally, when this model is placed on a graph, the ground state degeneracy is equal to the number of spanning trees of the graph, which is a common measure of complexity of the graph.

As explained in \cite{Gorantla:2022mrp}, the origin of the exponential behaviour of GSD is different from a similar phenomenon in a system of decoupled spins. Let us review it here for concreteness. In a system of decoupled spins, there is a symmetry generator associated with each spin, and all of them have the same order. So, on a 2d spatial torus lattice with $L_x = L_y = L$ sites in each direction, the number of generators is $L^2$, which leads to $\log \text{GSD} \sim L^2$. On the other hand, in the Laplacian $\phi$-theory, there are only $O(L)$ generators of the $\Jac(C_L \times C_L)$,\footnote{The minimal number of generators of $\Jac(C_{L_x}\times C_{L_y})$ is at most the number of nontrivial spatial integer gauge fields $n$'s after gauge fixing. One can gauge fix the $n$'s so that they are zero everywhere except at $x = 0,1$, or at $y = 0,1$. This means that the minimal number of generators of $\Jac(C_{L_x}\times C_{L_y})$ is at most $\min(2L_x,2L_y)$.} but some of them have very large orders, which leads to $\log \text{GSD} \sim L^2$.

Another consequence of this large symmetry is that the spatially separated two-point functions of monopole operators $e^{i\phi}$ and dipole operators $e^{i\Delta_i \phi}$ vanish. First, consider the spatially separated two-point function of the monopole operator
\ie
\langle e^{i\phi(\tau,x,y)} e^{-i\phi(0,0,0)} \rangle~,\qquad (x,y)\ne (0,0)~.
\fe
The discrete momentum symmetry of the 2+1d Laplacian $\phi$-theory includes linear shifts in $x$ and $y$:
\ie\label{xshift}
&\phi \rightarrow \phi + \frac{2\pi m_x x}{L_x} + \frac{2\pi m_y y}{L_y}~,
\\
&n \rightarrow n + m_x \left( \delta_{x,0} - \delta_{x,L_x-1} \right) + m_y \left( \delta_{y,0} - \delta_{y,L_y-1} \right)~,
\fe
where $m_i = 0,\ldots,L_i-1$. Under such shifts, the two-point function of the monopole operator acquires a nontrivial $(x,y)$-dependent phase, and hence it vanishes.

More interestingly, consider the spatially separated two-point function of the dipole operator
\ie\label{polyphi2:dip2pt}
\langle e^{i\Delta_x \phi(\tau,x+\frac12,y)} e^{-i\Delta_x \phi(0,\frac12,0)} \rangle~,\qquad (x,y)\ne (0,0)~.
\fe
It is clearly invariant under the linear shifts \eqref{xshift}. However, the discrete momentum symmetry of the 2+1d Laplacian $\phi$-theory also includes quadratic shifts in $x$ and $y$, such as
\ie\label{xyshift}
&\phi \rightarrow \phi + \frac{2\pi m x y}{\gcd(L_x,L_y)} + \frac{2\pi m' (x^2 - y^2 - L_x x + L_y y)}{2\gcd(L_x,L_y)}~,
\\
&n \rightarrow n + \frac{m}{\gcd(L_x,L_y)} \left[ L_x y(\delta_{x,0} - \delta_{x,L_x-1}) + L_y x (\delta_{y,0} - \delta_{y,L_y-1}) \right]
\\
&\qquad \quad - \frac{m'}{\gcd(L_x,L_y)} \left( L_x \delta_{x,0} - L_y \delta_{y,0} \right)~,
\fe
where $m = 0,\ldots,\gcd(L_x,L_y)$, and $m' = 0,\ldots, 2\gcd(L_x,L_y)$. Setting $L_x = L_y = L$ for simplicity, we see that the two-point function of the dipole operator is not invariant under these shifts, and hence it vanishes as well. Similar conclusion holds for the other dipole operator $e^{i\Delta_y \phi}$.

\subsubsection{Robustness}

We now discuss whether the GSD and the global symmetry of the low-energy limit of our modified Villain lattice is robust or not.
(See \cite{paper1} for general discussions on robustness and naturalness.)
More specifically, we impose the  momentum symmetry in the microscopic lattice model, and ask whether there are local, relevant operators in the low-energy limit that violate the winding symmetry and lift the GSD.
Typically, the notion of robustness requires a certain scaling symmetry to determine which operators are relevant and which ones are not.
Furthermore, the notion of a local operator only makes sense as we take the number of lattice sites to infinite, $L_i\to \infty$.
Our Laplacian $\phi$-theory does not admit a conventional continuum limit, so the above notions need to be appropriately generalized. Nonetheless, we will see that there are local operators (supported at a single site on the lattice) that lift the GSD and violate the winding symmetry. Since they act nontrivially on the space of ground states, they should be considered relevant in this sense.

Let us impose the $U(1)$ and discrete momentum symmetries. This symmetry excludes local operators such as $e^{i\Delta_i \phi}$, $e^{i\Delta_i \Delta_j \phi}$ in the Lagrangian.  In fact, this symmetry further excludes local operators of the form $\prod_{i=1}^n e^{iq_i \phi(0,x_i,y_i)}$, $q_i\in\mathbb Z$, that cannot be written as $\prod_{j=1}^m e^{ir_j (\Delta_x^2 + \Delta_y^2) \phi(0,x_j,y_j)}$, $r_j\in\mathbb Z$. We can think of these operators as being higher order than the operators that are already present in the action.  See Appendix \ref{app:polyphi} for more details.

However, we can add the winding operator $e^{i\tilde \phi}$ to the Lagrangian \eqref{2dlapphi-modVill-action} because it is invariant under the momentum symmetries. Such a perturbation breaks the winding symmetry and hence it lifts the ground states. Therefore, if we impose only the momentum symmetries, the ground state degeneracy is not robust.

\subsection{Dipole $\phi$-theory}\label{sec:2ddipphi}
In this subsection, we study the modified Villain model associated with the second discretization of $(\nabla^2\phi)^2$. We refer to it as the 2+1d dipole $\phi$-theory.

Let us place the theory on a periodic 3d Euclidean spacetime lattice with $L_\tau$, $L_x$, and $L_y$ sites. The modified Villain action of the 2+1d dipole $\phi$-theory is
\ie\label{2ddipphi-modVill-action}
S &= \frac{\beta_0}{2} \sum_{\tau\text{-link}} (\Delta_\tau \phi - 2\pi n_\tau)^2 + \frac{\beta'}{2} \sum_\text{site} \left[(\Delta_x^2 \phi - 2\pi n_{xx})^2 + (\Delta_y^2 \phi - 2\pi n_{yy})^2\right]
\\
& + \frac{\beta}{2} \sum_{xy\text{-plaq}} (\Delta_x\Delta_y \phi - 2\pi n_{xy})^2 + i\sum_{x\text{-link}} \hat{\mathcal A}_{\tau x} (\Delta_x n_{yy} - \Delta_y n_{xy}) + i\sum_{y\text{-link}} \hat{\mathcal A}_{\tau y} (\Delta_y n_{xx} - \Delta_x n_{xy})
\\
& + i \sum_\text{cube} \hat{\mathcal A}_{xy} (\Delta_\tau n_{xy} - \Delta_x \Delta_y n_\tau) - i \sum_{\tau\text{-link}} \hat{\mathcal A}_{yy} (\Delta_\tau n_{xx} - \Delta_x^2 n_\tau) - i \sum_{\tau\text{-link}} \hat{\mathcal A}_{xx} (\Delta_\tau n_{yy} - \Delta_y^2 n_\tau)~,
\fe
where $\beta$ and $\beta'$ are not \emph{a priori} related to each other, whereas $\beta = 2\beta'$ when related to \eqref{2dphi-second-discretization}. Here, $\phi$ is a real-valued scalar field, $(n_\tau, n_{ij})$ are integer gauge fields, and $(\hat{\mathcal A}_{\tau i},\hat{\mathcal A}_{ij})$ are real-valued Lagrange multipliers that make the integer gauge fields flat. The locations of these fields on the lattice are summarized in Table \ref{tbl:2dphiloc}. They have a gauge symmetry
\ie\label{2ddipphi-modVill-gaugesym}
&\phi \sim \phi + 2\pi k~, &&\hat{\mathcal A}_{\tau i} \sim \hat{\mathcal A}_{\tau i} + \Delta_\tau \hat \alpha_i + 2\pi \hat k_{\tau i}~,
\\
&n_\tau \sim n_\tau + \Delta_\tau k~, && \hat{\mathcal A}_{ii} \sim \hat{\mathcal A}_{ii} + \Delta_i \hat \alpha_i + 2\pi \hat k_{ii}~,
\\
&n_{ij} \sim n_{ij} + \Delta_i \Delta_j k~, \qquad &&\hat{\mathcal A}_{xy} \sim \hat{\mathcal A}_{xy} + \Delta_x \hat \alpha_y + \Delta_y \hat \alpha_x + 2\pi \hat k_{xy}~,
\fe
where $k$ and $\hat k$'s are integer gauge parameters, and $\hat \alpha_i$ is a real-valued gauge parameter along the $\tau i$-plaquette. The integer gauge symmetry makes $\phi$ and $(\hat{\mathcal A}_{\tau i},\hat{\mathcal A}_{ij})$ compact.

\subsubsection{Lineon-elasticity duality}\label{sec:lineonelasticity}
Using the Poisson resummation formula for the integers $(n_\tau,n_{ij})$,  the action \eqref{2ddipphi-modVill-action} is dualized to
\ie\label{2ddipphi-modVill-dual-action}
S &= \frac{\hat \gamma_0}{2} \sum_{xy\text{-plaq}} \hat{\mathcal E}_{xy}^2 + \frac{\hat \gamma'_0}{2} \sum_\text{site} ( \hat{\mathcal E}_{xx}^2 + \hat{\mathcal E}_{yy}^2 ) + \frac{\hat \gamma}{2} \sum_{\tau\text{-link}} \hat{\mathcal B}^2
\\
&\qquad - i \sum_\text{site} \phi (\Delta_\tau \hat n - \Delta_x \Delta_y \hat n_{\tau xy} + \Delta_x^2 \hat n_{\tau yy} + \Delta_y^2 \hat n_{\tau xx})~,
\fe
where
\ie
\hat \gamma_0 = \frac{1}{(2\pi)^2\beta}~,\qquad \hat \gamma'_0 = \frac{1}{(2\pi)^2\beta'}~,\qquad \hat \gamma = \frac{1}{(2\pi)^2\beta_0}~,
\fe
and the field strengths are
\ie
&\hat{\mathcal E}_{xy} = \Delta_\tau \hat{\mathcal A}_{xy} - \Delta_x \hat{\mathcal A}_{\tau y} - \Delta_y \hat{\mathcal A}_{\tau x} - 2\pi \hat n_{\tau xy}~,
\\
&\hat{\mathcal E}_{ii} = \Delta_\tau \hat{\mathcal A}_{ii} - \Delta_i \hat{\mathcal A}_{\tau i} - 2\pi \hat n_{\tau ii}~,
\\
&\hat{\mathcal B} = \Delta_x \Delta_y \hat{\mathcal A}_{xy} - \Delta_x^2 \hat{\mathcal A}_{yy} - \Delta_y^2 \hat{\mathcal A}_{xx} - 2\pi \hat n~.
\fe
The integer gauge fields $(\hat n_{\tau ij},\hat n)$ have the gauge symmetry
\ie
&\hat n_{\tau xy} \sim \hat n_{\tau xy} + \Delta_\tau \hat k_{xy} - \Delta_x \hat k_{\tau y} - \Delta_y \hat k_{\tau x}~,
\\
&\hat n_{\tau ii} \sim \hat n_{\tau ii} + \Delta_\tau \hat k_{ii} - \Delta_i \hat k_{\tau i}~,
\\
&\hat n \sim \hat n + \Delta_x \Delta_y \hat k_{xy} - \Delta_x^2 \hat k_{yy} - \Delta_y^2 \hat k_{xx}~.
\fe
This modified Villain model is a lattice discretization of the 2+1d vector charge theory of \cite{Xu:2006,Bulmash:2018lid,Oh:2021gee,Oh:2022klh} in the continuum (see also \cite{rasmussen,Pretko:2016kxt,Pretko:2016lgv,Pretko:2017xar,Pretko:2018jbi} for a similar gauge theory in 3+1d).  See Table \ref{tbl:contlag} for information about the continuum theory.

In the continuum, the duality is schematically given by the map,
\ie
\partial_\tau \phi &\leftrightsquigarrow \hat B = \partial_x \partial_y \hat A_{xy} - \partial_x^2 \hat A_{yy} - \partial_y^2 \hat A_{xx}~,
\\
\partial_x \partial_y \phi &\leftrightsquigarrow \hat E_{xy} = \partial_\tau \hat A_{xy} - \partial_x \hat A_{\tau y} - \partial_y \hat A_{\tau x}~,
\\
\partial_i^2 \phi &\leftrightsquigarrow \hat E_{jj} = \partial_\tau \hat A_{jj} - \partial_j \hat A_{\tau j}~,\qquad i,j=x,y,\quad i\ne j~,
\fe
where $(\hat A_{\tau i}, \hat A_{ij})$ are gauge fields with gauge transformations
\ie
&\hat A_{\tau i} \sim \hat A_{\tau i} + \partial_\tau \hat \alpha_i~,\quad \hat A_{xy} \sim \hat A_{xy} + \partial_x \hat \alpha_y + \partial_y \hat \alpha_x~,\quad \hat A_{ii} \sim \hat A_{ii} + \partial_i \hat \alpha_i~,
\fe
and $\hat \alpha_i$ are gauge parameters.

The continuum version of the duality between the dipole $\phi$-theory and the vector charge theory was discussed in \cite{Manoj:2020bcz} in the context of elasticity theory. Since this model has defects representing the worldline of lineons (see Section \ref{sec:2ddipphi-sym} below), we dub this duality as the lineon-elasticity duality.

\subsubsection{Global symmetry, ground state degeneracy and mobility of defects}\label{sec:2ddipphi-sym}
Let us discuss the global symmetry of the 2+1d dipole $\phi$-theory \eqref{2ddipphi-modVill-action}. (See Appendix \ref{app:2ddipphi} for more details.)
\begin{itemize}
\item The $U(1)$ momentum (dual magnetic) symmetry acts as $\phi \rightarrow \phi + c$, where $c\sim c+2\pi$ is a circle-valued constant. It is a magnetic global symmetry from the dual gauge theory of point of view.

\item The $\mathbb Z_{L_x}$ momentum (dual magnetic) dipole symmetry acts as
\ie\label{2ddipphi-linear-shift}
&\phi \rightarrow \phi + 2\pi m_{xx} \frac{x}{L_x}~,
\\
&n_{xx} \rightarrow n_{xx} + m_{xx} \left( \delta_{x,0} - \delta_{x,L_x-1} \right)~,
\fe
where $m_{xx}=0,1,\ldots, L_x-1$. Note that the $m_{xx}=L_x$ transformation is trivial because it can be undone by an integer gauge transformation \eqref{2ddipphi-modVill-gaugesym} with $k=x$. More intuitively, the $m_{xx}=L_x$ transformation is trivial because $\phi$ is compact. Also note that this transformation is compatible with the periodicity of the spatial torus lattice, i.e., $x\sim x+L_x$.

\item The $\mathbb Z_{L_y}$ momentum (dual magnetic) dipole symmetry acts similarly with $x$ and $y$ exchanged.

\item The $U(1)^3$ winding (dual electric) symmetry acts as
\ie
\hat{\mathcal A}_{ii} \rightarrow \hat{\mathcal A}_{ii} + \frac{\hat c_{ii}}{L_i}~,\qquad \hat{\mathcal A}_{xy} \rightarrow \hat{\mathcal A}_{xy} + \frac{\hat c_{xy}}{\lcm(L_x,L_y)}~,
\fe
where $\hat c_{ij}\sim \hat c_{ij} + 2\pi$ are circle-valued constants.\footnote{In particular, the transformations with $\hat c_{ij}=2\pi$ are trivial and can be undone by gauge transformations of the form \eqref{2ddipphi-modVill-gaugesym}. Similar comments apply to various other electric global symmetries discussed below. See \cite{Gorantla:2022eem} for a simpler example in 1+1d.}

\item The $\mathbb Z_{L_x}$ winding (dual electric) dipole symmetry acts as
\ie\label{dipolephi:windingdipole}
&\hat{\mathcal A}_{yy} \rightarrow \hat{\mathcal A}_{yy} + 2\pi \hat m_{xx} \frac{x}{L_x} \delta_{y,0}~,
\\
&\hat n \rightarrow \hat n - \hat m_{xx} \delta_{y,0} (\delta_{x,0} - \delta_{x,L_x-1})~.
\fe
where $\hat m_{xx} = 0,\ldots,L_x-1$.

\item The $\mathbb Z_{L_y}$ winding (dual electric) dipole symmetry acts similarly with $x$ and $y$ exchanged.
\end{itemize}
The momentum and winding dipole symmetries do not commute with each other, which leads to a ground state degeneracy of $L_x L_y$. See the discussion around \eqref{phinoncommute} for more details. In fact, every state in the Hilbert space is $L_x L_y$-fold degenerate.  More abstractly, the momentum and winding dipole symmetries are realized projectively.  This can be viewed as a mixed anomaly between them.

Let us briefly discuss the fate of these symmetries in a specific continuum limit. (Recall that, as in the 1+1d version of this theory \cite{Gorantla:2022eem}, there are several distinct continuum limits.) We introduce the lattice spacings $a_\tau,a$ and take the limit $a_\tau,a\rightarrow 0$ and $L_i \rightarrow \infty$ with $\ell_\tau = a_\tau L_\tau$ and $\ell_i = a L_i$ fixed, while scaling the lattice coupling constants as
\ie
\beta_0 = \frac{\mu_0 a^2}{a_\tau}~,\qquad \beta = 2\beta' = \frac{2a_\tau}{\mu a^2}~,
\fe
where $\mu,\mu_0$ are fixed continuum coupling constants with mass dimensions $1$. In this continuum limit, the above global symmetries reduce to the global symmetries of the 2+1d continuum compact Lifshitz theory discussed at the beginning of Section \ref{sec:2dphi}:\footnote{In taking the continuum limit of the $\mathbb Z_{\gcd(L_x,L_y)}$ time-like winding dipole symmetry, we assume that $L_x/L_y$ is fixed.}
\ie
\text{Space-like symmetries:}\quad\qquad\qquad&
\\
U(1)\text{ momentum} \quad &\longrightarrow \quad U(1)\text{ momentum},
\\
\mathbb Z_{L_x} \times \mathbb Z_{L_y}\text{ momentum dipole} \quad &\longrightarrow \quad \mathbb Z\times \mathbb Z\text{ momentum dipole},
\\
U(1)^3\text{ winding} \quad &\longrightarrow \quad \text{does not exist},
\\
\mathbb Z_{L_x} \times \mathbb Z_{L_y}\text{ winding dipole} \quad &\longrightarrow \quad \text{part of }U(1)\text{ ``winding dipole'' one-form}.
\\
\text{Time-like symmetries:}\quad\qquad\qquad&
\\
U(1)^2\text{ winding} \quad &\longrightarrow \quad \text{does not exist},
\\
\mathbb Z_{\gcd(L_x,L_y)}\text{ winding dipole} \quad &\longrightarrow \quad \text{part of }U(1)\text{ ``winding dipole'' one-form}.
\fe
In particular, the two integer-valued charges \eqref{2ddipphi-U1wind1form} of the $U(1)$ ``winding dipole'' one-form symmetry generate two space-like $U(1)$ symmetries that are the continuum limits of the $\mathbb Z_{L_x}$ and $\mathbb Z_{L_y}$ winding dipole symmetries on the lattice.

Note that the $\mathbb Z_{L_x} \times \mathbb Z_{L_y}$ momentum dipole symmetry of the 2+1d dipole $\phi$-theory is smaller than the discrete momentum symmetry of the 2+1d Laplacian $\phi$-theory (see Section \ref{sec:2dlapphi-sym}). For example, in addition to the linear shifts of \eqref{2ddipphi-linear-shift} (or equivalently, \eqref{xshift}), \eqref{2dlapphi-mom-har} includes the quadratic shifts \eqref{xyshift}. Consequently, while the two-point function of the monopole operator $e^{i\phi}$ vanishes in both Laplacian and dipole $\phi$-theories, the two-point function of the dipole operator $e^{i\Delta_i \phi}$ vanishes only in the Laplacian $\phi$-theory.

There are also time-like global winding symmetries that act on defects that extend in the Euclidean time direction such as
\ie\label{dipolephi:hatAdefect}
\exp\left( i \sum_{x\text{-link: fixed }x,y} \hat{\mathcal A}_{\tau x} \right)~,
\fe
which represents the worldline of a static particle. We have:
\begin{itemize}
\item The $U(1)^2$ time-like winding (dual electric) symmetry acts as
\ie
\hat{\mathcal A}_{\tau i} \rightarrow \hat{\mathcal A}_{\tau i} + \frac{\hat c_{\tau i}}{L_\tau}~,
\fe
where $\hat c_{\tau i} \sim \hat c_{\tau i} + 2\pi$ are circle-valued constants. The charge, which is commonly referred to as the ``gauge charge'', associated with this time-like global symmetry generated by $\hat c_{\tau i}$ is a vector in space, and hence the name vector charge theory.

\item The $\mathbb Z_{\gcd(L_x,L_y)}$ time-like winding (dual electric) dipole symmetry acts as
\ie
&\hat{\mathcal A}_{\tau x}(\tau,x+\tfrac12,y) \rightarrow \hat{\mathcal A}_{\tau x}(\tau,x+\tfrac12,y) - 2\pi \hat m_{\tau xy} \delta_{\tau,0} \frac{y}{\gcd(L_x,L_y)}~,
\\
&\hat{\mathcal A}_{\tau y}(\tau,x,y+\tfrac12) \rightarrow \hat{\mathcal A}_{\tau y}(\tau,x,y+\tfrac12) + 2\pi \hat m_{\tau xy} \delta_{\tau,0} \frac{x}{\gcd(L_x,L_y)}~,
\\
&\hat n_{\tau xy}(\tau,x+\tfrac12,y+\tfrac12) \rightarrow \hat n_{\tau xy}(\tau,x+\tfrac12,y+\tfrac12) + \frac{\hat m_{\tau xy} \delta_{\tau,0}}{\gcd(L_x,L_y)} \left( L_x \delta_{x,L_x-1} - L_y \delta_{y,L_y-1} \right)~,
\fe
where $\hat m_{\tau xy} = 0,\ldots,\gcd(L_x,L_y)-1$.
\end{itemize}
As a consequence of the time-like dipole symmetry, the particle associated with the static defect \eqref{dipolephi:hatAdefect} can move in the $x$-direction, but it can only hop by $\gcd(L_x,L_y)$ sites in the $y$-direction. Hence, we call it an $x$-lineon. Similarly, there is a $y$-lineon associated with the defect of $\hat{\mathcal A}_{\tau y}$. Therefore, the dipole $\phi$-theory is a model of lineons.

\subsubsection{Robustness}
Let us impose the $U(1)$ momentum and $\mathbb Z_{L_x}\times \mathbb Z_{L_y}$ momentum dipole symmetries. In this case, there are no lower order difference terms that one can add to the action \eqref{2ddipphi-modVill-action} because $e^{i\Delta_x \phi}$ and $e^{i\Delta_y \phi}$ are not invariant under the momentum dipole symmetry. In other words, the action \eqref{2ddipphi-modVill-action} includes the most relevant terms that preserve the momentum symmetries. Moreover, the winding symmetries are robust because the operators charged under them are extended in space. In particular, the $L_x L_y$-fold ground state degeneracy is robust once we impose the momentum symmetries.

Alternatively, we can start with the dual gauge theory and impose the dual electric (winding) symmetries. In this case, the ground state degeneracy is not robust because we can always perturb the dual action \eqref{2ddipphi-modVill-dual-action} by the monopole operator $e^{i\phi}$ which acts nontrivially on the ground states.

This situation is reminiscent of the robustness of an ordinary $U(1)$ gauge theory in 2+1d or its dual compact scalar theory.  Imposing the momentum symmetry of the scalar, which is the magnetic zero-form symmetry of its dual gauge theory, makes the theory robust.  Its other symmetry is a winding one-form symmetry, or its dual a one-form electric symmetry, but there is no local operator charged under it.  Conversely, imposing only the latter symmetry, the theory is not robust.  Now, there are local operators that are charged under the momentum symmetry of the scalar or equivalently, the magnetic symmetry of the gauge theory.  These operators can be present in the Lagrangian and gap the system.  This is the famous Polyakov mechanism \cite{POLYAKOV1977429}.

\section{2+1d $U(1)$ tensor gauge theories}\label{sec:2dA}

Since the momentum symmetries of the 2+1d dipole and Laplacian $\phi$-theories are very different, the corresponding $U(1)$ gauge theories are also very different.  These are, respectively, the scalar charge theory and the Laplacian gauge theory in Table \ref{tbl:contlag}. In this section, we discuss the modified Villain formulation of the two $U(1)$ tensor gauge theories. The differences between these two theories are summarized in Table \ref{tbl:2dA}.

\begin{table}
\begin{center}
\begin{tabular}{|c|c|c|}
\hline
\multirow{2}{*}{Theory} & Laplacian gauge theory & Scalar charge theory
\tabularnewline
& $({\cal A}_\tau ,{\cal A})$ &$({\cal A}_\tau, {\cal A}_{ij})$
\tabularnewline
\hline
Duality & No duality & Dual to the matter theory $\hat\phi_i$
\tabularnewline
\hline
\multirow{2}{*}{Space-like global symmetry} & \multirow{2}{*}{$U(1)$ electric} & $U(1)^3 \times \mathbb Z_{\gcd(L_x,L_y)}$ electric
\tabularnewline
& & $U(1)^2 \times \mathbb Z_{\gcd(L_x,L_y)}$ magnetic
\tabularnewline
\hline
Time-like global symmetry & $U(1)\times \Jac$ & $U(1)\times \mathbb Z_{L_x} \times \mathbb Z_{L_y}$
\tabularnewline
\hline
Ground state degeneracy & 1 & $\gcd(L_x,L_y)$
\tabularnewline
\hline
Robustness & Yes & No
\tabularnewline
\hline
\end{tabular}
\end{center}
\caption{Comparison of the two $U(1)$ tensor gauge theories associated with the momentum symmetries of the two models in Section \ref{sec:2dphi}. Here, ``$\Jac$'' is short-hand for the Jacobian group $\Jac(C_{L_x} \times C_{L_y})$ of the 2d torus graph $C_{L_x} \times C_{L_y}$. The last row refers to robustness of the magnetic symmetry and ground state degeneracy after imposing the electric symmetry. Since the Laplacian gauge theory does not have a magnetic symmetry or nontrivial ground state degeneracy, it is trivially robust.}\label{tbl:2dA}
\end{table}

\subsection{Laplacian gauge theory}\label{sec:2dlapA}
We can gauge the momentum symmetry of the Laplacian $\phi$-theory of Section \ref{sec:2dlapphi} by coupling it to the gauge fields $(\mathcal A_\tau,\mathcal A;m_\tau)$, where ${\cal A}_\tau,{\cal A}$ are real-valued and $m_\tau$ is integer-valued. In this section, we study the pure gauge theory of $(\mathcal A_\tau,\mathcal A;m_\tau)$ described by the action
\ie\label{2dlapA-modVill-action}
S = \frac{\gamma}{2} \sum_{\tau\text{-link}} \mathcal E^2~,
\fe
where $\mathcal E = \Delta_\tau \mathcal A - (\Delta_x^2 + \Delta_y^2) \mathcal A_\tau - 2\pi m_\tau$ is the electric field of $(\mathcal A_\tau,\mathcal A;m_\tau)$.  It is invariant under the gauge symmetry
\ie
&\mathcal A_\tau \sim \mathcal A_\tau + \Delta_\tau \alpha + 2\pi q_\tau~,
\\
&\mathcal A \sim \mathcal A + (\Delta_x^2 + \Delta_y^2) \alpha + 2\pi q~,
\\
&m_\tau \sim m_\tau + \Delta_\tau q - (\Delta_x^2 + \Delta_y^2) q_\tau~,
\fe
where $(q_\tau,q)$ are integer gauge parameters, and $\alpha$ is a real gauge parameter. The locations of these fields on the lattice are summarized in Table \ref{tbl:2dAloc}. This integer gauge symmetry makes the gauge fields $(\mathcal A_\tau,\mathcal A)$ compact.

\begin{table}
\begin{center}
\begin{tabular}{|c|c|c|}
\hline
Location & Laplacian gauge theory & Scalar charge theory
\tabularnewline
\hline
site & $\mathcal A$ & $\mathcal A_{ii}$
\tabularnewline
\hline
$\tau$-link & $\mathcal A_{\tau},m_\tau$ & $\mathcal A_{\tau},n_{\tau ii},\hat n_{ii}$
\tabularnewline
\hline
$i$-link & -- & $n_i,\hat n_{\tau i}$
\tabularnewline
\hline
$\tau i$-plaq & -- & $\hat \phi_i$
\tabularnewline
\hline
$xy$-plaq & -- & $\mathcal A_{xy}$
\tabularnewline
\hline
cube & -- & $n_{\tau xy},\hat n_{xy}$
\tabularnewline
\hline
\end{tabular}
\end{center}
\caption{Locations on the lattice of various fields of the Laplacian and the scalar charge theories of Section \ref{sec:2dA}. Here, $i,j=x,y$, $i\ne j$, and repeated indices are not summed over.}\label{tbl:2dAloc}
\end{table}

We can add a $\theta$-term to the action \eqref{2dlapA-modVill-action}:
\ie
\frac{i\theta}{2\pi} \sum_{\tau\text{-link}} \mathcal E~.
\fe
Here, $\theta \sim \theta + 2\pi$ because $\sum_{\tau\text{-link}} \mathcal E = -2\pi \sum_{\tau\text{-link}} m_\tau \in 2\pi \mathbb Z$.

Furthermore, this theory is robust because all the local operators are polynomials in the electric field strength and it lattice derivatives.  Hence, they are higher order than the terms in the action.

As in Section \ref{sec:2dlapphi}, the 2+1d $U(1)$ Laplacian gauge theory can be placed on a general spatial graph \cite{Gorantla:2022mrp}.

\subsubsection{Global symmetry and mobility of defects}
Let us discuss the space-like and time-like global symmetries of the 2+1d $U(1)$ Laplacian gauge theory.
\begin{itemize}
\item There is a $U(1)$ electric space-like symmetry that acts as $\mathcal A \rightarrow \mathcal A + \frac{c}{L_xL_y}$, where $c \sim c+2\pi$ is a circle-valued constant. The charged operator is a surface operator supported on the whole space
\ie
\exp\left(i \sum_{\text{site: fixed }\tau} \mathcal A\right)~.
\fe

\item There is a $U(1)$ electric time-like symmetry that acts as $\mathcal A_\tau \rightarrow \mathcal A_\tau + \delta_{\tau,0} c_\tau$, where $c_\tau \sim c_\tau + 2\pi$ is a circle-valued constant. The charged object is a line defect that extends in the Euclidean time direction
\ie\label{2dlapA-defect}
\exp\left(i \sum_{\tau\text{-link: fixed }x,y} \mathcal A_\tau\right)~.
\fe

\item There is a discrete electric time-like symmetry that acts as
\ie
&\mathcal A_\tau(\tau+\tfrac12,x,y) \rightarrow \mathcal A_\tau(\tau+\tfrac12,x,y) + \delta_{\tau,0} f_\tau(x,y)~,
\\
&m_\tau(\tau+\tfrac12,x,y) \rightarrow m_\tau(\tau+\tfrac12,x,y) - \frac{1}{2\pi}\delta_{\tau,0}(\Delta_x^2+\Delta_y^2)f_\tau(x,y)~,
\fe
where $f_\tau(x,y)$ is a circle-valued discrete harmonic function on the 2d spatial torus lattice. The corresponding symmetry group is the Jacobian group, $\text{Jac}(C_{L_x}\times C_{L_y})$, of the 2d torus lattice.
\end{itemize}
It was proven in \cite{Gorantla:2022mrp} that the discrete electric time-like symmetry implies that the particle described by the static defect \eqref{2dlapA-defect} is immobile, i.e., it is a fracton. In fact, if we place the theory on an infinite square lattice $\mathbb Z^2$, then we show, in Appendix \ref{app:polyA}, that any finite set of particles is immobile unless they are in the trivial superselection sector, i.e., they can be ``annihilated locally.''

\subsection{Scalar charge theory}\label{sec:2ddipA}
We can gauge the momentum symmetry of the 2+1d dipole $\phi$-theory of Section \ref{sec:2ddipphi} by coupling it to the rank-2 $U(1)$ tensor gauge fields $(\mathcal A_\tau, \mathcal A_{ij};n_{\tau ij},n_x,n_y)$. In this section, we will study the pure gauge theory of these gauge fields. Our lattice model is a discretization of the continuum scalar charge theory of \cite{Pretko:2016kxt,Pretko:2016lgv,Pretko:2018qru,Bulmash:2018lid,Slagle:2018kqf,Pretko:2019omh,Shenoy:2019wng,Oh:2021gee,Oh:2022klh}. The action is
\ie\label{2ddipA-modVill-action}
S &= \frac{\gamma_0}{2} \sum_\text{cube} \mathcal E_{xy}^2 + \frac{\gamma'_0}{2} \sum_{\tau\text{-link}} (\mathcal E_{xx}^2 + \mathcal E_{yy}^2) + \frac{\gamma}{2} \left( \sum_{x\text{-link}} \mathcal B_x^2 + \sum_{y\text{-link}} \mathcal B_y^2 \right)
\\
& + i \sum_{\tau x\text{-plaq}} \hat \phi_x (\Delta_\tau n_x - \Delta_x n_{\tau yy} + \Delta_y n_{\tau xy}) + i \sum_{\tau y\text{-plaq}} \hat \phi_y (\Delta_\tau n_y - \Delta_y n_{\tau xx} + \Delta_x n_{\tau xy})~,
\fe
where
\ie\label{2ddipA-fieldstrengths}
&\mathcal E_{ij} = \Delta_\tau \mathcal A_{ij} - \Delta_i \Delta_j \mathcal A_\tau - 2\pi n_{\tau ij}~,
\\
&\mathcal B_x = \Delta_x \mathcal A_{yy} - \Delta_y \mathcal A_{xy} - 2\pi n_x~,
\\
&\mathcal B_y = \Delta_y \mathcal A_{xx} - \Delta_x \mathcal A_{xy} - 2\pi n_y~,
\fe
are the gauge invariant field strengths of $(\mathcal A_\tau,\mathcal A_{ij};n_{\tau ij},n_i)$. Here, $(\mathcal A_\tau, \mathcal A_{ij})$ are real gauge fields, $(n_{\tau ij},n_i)$ are integer gauge fields, and $\hat \phi_i$ are real Lagrange multipliers that make the integer gauge fields flat. The locations on the lattice of these fields are summarized in Table \ref{tbl:2dAloc}. The gauge symmetry is
\ie
&\mathcal A_\tau \sim \mathcal A_\tau + \Delta_\tau \alpha + 2\pi k_\tau~,\qquad && n_{\tau ij} \sim n_{\tau ij} + \Delta_\tau k_{ij} - \Delta_i \Delta_j k_\tau~,
\\
&\mathcal A_{ij} \sim \mathcal A_{ij} + \Delta_i \Delta_j \alpha + 2\pi k_{ij}~,\qquad && n_x \sim n_x + \Delta_x k_{yy} - \Delta_y k_{xy}~,
\\
&\hat \phi_i \sim \hat \phi_i + 2\pi \hat k_i~,\qquad && n_y \sim n_y + \Delta_y k_{xx} - \Delta_x k_{xy}~,
\fe
where $\alpha$ is a real gauge parameter, and $(k_\tau,k_{ij})$ and $\hat k_i$ are integer gauge parameters.

\subsubsection{Fracton-elasticity duality}\label{sec:fractonelasticity}
Using the Poisson resummation formula for the integers $(n_{\tau ij}, n_x,n_y)$, the action \eqref{2ddipA-modVill-action} is dualized to
\ie\label{2ddipA-modVill-dual-action}
S &= \frac{\hat \beta_0}{2} \left[ \sum_{x\text{-link}} (\Delta_\tau \hat \phi_x - 2\pi \hat n_{\tau x})^2 + \sum_{y\text{-link}} (\Delta_\tau \hat \phi_y - 2\pi \hat n_{\tau y})^2 \right] + \frac{\hat \beta}{2} \sum_\text{cube} (\Delta_x \hat \phi_y + \Delta_y \hat \phi_x - 2\pi \hat n_{xy})^2
\\
& + \frac{\hat \beta'}{2} \sum_{\tau\text{-link}} \sum_i (\Delta_i \hat \phi_i - 2\pi \hat n_{ii})^2 - i \sum_{\tau\text{-link}} \mathcal A_\tau (\Delta_x \Delta_y \hat n_{xy} - \Delta_x^2 \hat n_{yy} - \Delta_y^2 \hat n_{xx})
\\
&+ i \sum_\text{site} \sum_{i\ne j} \mathcal A_{ii} (\Delta_\tau \hat n_{jj} - \Delta_j \hat n_{\tau j}) - i \sum_{xy\text{-plaq}} \mathcal A_{xy} (\Delta_\tau \hat n_{xy} - \Delta_x \hat n_{\tau y} - \Delta_y \hat n_{\tau x})~,
\fe
where
\ie
\hat \beta_0 = \frac{1}{(2\pi)^2 \gamma}~,\qquad \hat \beta = \frac{1}{(2\pi)^2 \gamma_0}~,\qquad \hat \beta' = \frac{1}{(2\pi)^2 \gamma'_0}~,
\fe
and the integer gauge fields $(\hat n_{\tau i},\hat n_{ij})$ have a gauge symmetry
\ie
\hat n_{\tau i} \sim \hat n_{\tau i} + \Delta_\tau \hat k_i~,\qquad \hat n_{ii} \sim \hat n_{ii} + \Delta_i \hat k_i~,\qquad \hat n_{xy} \sim \hat n_{xy} + \Delta_x \hat k_y + \Delta_y \hat k_x~.
\fe
The theory \eqref{2ddipA-modVill-dual-action} can be thought of as a ``matter theory'' with the matter fields $(\hat \phi_x, \hat \phi_y)$.  It  is closely related to the gauge theory \eqref{2ddipphi-modVill-dual-action}.  Gauging the momentum symmetry of the matter theory \eqref{2ddipA-modVill-dual-action} couples the theory to the gauge fields $(\hat{\mathcal A}_{\tau i},\hat{\mathcal A}_{ij})$ of \eqref{2ddipphi-modVill-dual-action}. The relations between the dipole $\phi$-theory, the vector charge theory $(\hat{\cal A}_{\tau i}, \hat {\cal A}_{ij})$, the scalar charge theory $({\cal A}_\tau , {\cal A}_{ij})$, and the matter theory $\hat \phi_i$ are summarized in Figure \ref{fig:duality}.\footnote{Their relations are similar to the four theories (with similar notations) studied in \cite{paper2}.}

The continuum Lagrangian for the matter fields $(\hat\phi_x,\hat\phi_y)$ takes the form
\ie
\mathcal L = \frac{\hat \mu_0}{2} \sum_i (\partial_\tau \hat \phi_i)^2 + \frac{\hat \mu}{2} \sum_{i,j} (\partial_i \hat \phi_j + \partial_j \hat \phi_i)^2~.
\fe
The fields $(\hat\phi_x,\hat\phi_y)$ have figured as displacement fields in an effective field theory for elasticity, and its relation to the scalar charge theory was explored in \cite{Pretko:2018qru,Pretko:2019omh,Nguyen:2020yve}. Specifically, disclinations in the elasticity theory correspond to fracton defects in the scalar charge theory. See also \cite{Chen:2020jew} for another appearance of this theory. The duality is schematically given by the map,
\ie
\partial_\tau \hat \phi_i &\leftrightsquigarrow B_i = \partial_i A_{jj} - \partial_j A_{ij}~,\qquad i,j=x,y,\quad i\ne j~,
\\
\partial_i \hat \phi_i &\leftrightsquigarrow E_{jj} = \partial_\tau A_{jj} - \partial_j^2 A_\tau~,\qquad i,j=x,y,\quad i\ne j~,
\\
\partial_x \hat \phi_y + \partial_y \hat \phi_x &\leftrightsquigarrow E_{xy} = \partial_\tau A_{xy} - \partial_x \partial_y A_\tau~.
\fe
Our exact duality above is a lattice version of the fracton-elasticity duality in the continuum.

\subsubsection{Global symmetry, ground state degeneracy, and mobility of defects}
Let us discuss the global symmetry of the modified Villain version of the 2+1d scalar charge theory. (See Appendix \ref{app:2ddipA} for more details.)
\begin{itemize}
\item The $U(1)^3$ electric (dual winding) symmetry acts as
\ie
\mathcal A_{ii} \rightarrow \mathcal A_{ii} + \frac{c_{ii}}{L_i}~,\qquad \mathcal A_{xy} \rightarrow \mathcal A_{xy} + \frac{c_{xy}}{\gcd(L_x,L_y)}~,
\fe
where $c_{ij} \sim c_{ij} + 2\pi$ are circle-valued constants.

\item The $\mathbb Z_{\gcd(L_x,L_y)}$ electric (dual winding) dipole symmetry acts as
\ie\label{2ddipA-elecgcd}
&\mathcal A_{xy}(\tau,x+\tfrac12,y+\tfrac12) \rightarrow \mathcal A_{xy}(\tau,x+\tfrac12,y+\tfrac12) + \frac{2\pi m_{xy}}{\gcd(L_x,L_y)} \left(\tilde L_y \delta_{x,L_x-1} - \tilde L_x \delta_{y,L_y-1}\right)~,
\\
&\mathcal A_{xx}(\tau,x,y) \rightarrow \mathcal A_{xx}(\tau,x,y) - \frac{2\pi m_{xy}}{\gcd(L_x,L_y)} \tilde L_y y (\delta_{x,0} - \delta_{x,L_x-1})~,
\\
&\mathcal A_{yy}(\tau,x,y) \rightarrow \mathcal A_{yy}(\tau,x,y) + \frac{2\pi m_{xy}}{\gcd(L_x,L_y)} \tilde L_x x (\delta_{y,0} - \delta_{y,L_y-1})~,
\\
&n_x(\tau,x+\tfrac12,y) \rightarrow n_x(\tau,x+\tfrac12,y) - \frac{m_{xy}}{\gcd(L_x,L_y)} \tilde L_x L_x (\delta_{y,0} - \delta_{y,L_y-1}) \delta_{x,L_x-1}~,
\\
&n_y(\tau,x,y+\tfrac12) \rightarrow n_y(\tau,x,y+\tfrac12) + \frac{m_{xy}}{\gcd(L_x,L_y)} \tilde L_y L_y (\delta_{x,0} - \delta_{x,L_x-1}) \delta_{y,L_y-1}~,
\fe
where $m_{xy} = 0,\ldots,\gcd(L_x,L_y)-1$, and $\tilde L_i$ are integer solutions of the equation $\tilde L_x L_x + \tilde L_y L_y = \gcd(L_x,L_y)$.

\item The $U(1)^2$ magnetic (dual momentum) symmetry acts as $\hat \phi_i \rightarrow \hat \phi_i + \hat c_i$, where $\hat c_i \sim \hat c_i + 2\pi$ are circle-valued constants.

\item The $\mathbb Z_{\gcd(L_x,L_y)}$ magnetic (dual momentum) dipole symmetry acts as
\ie\label{2ddipA-maggcd}
&\hat \phi_x(\tau+\tfrac12,x+\tfrac12,y) \rightarrow \hat \phi_x(\tau+\tfrac12,x+\tfrac12,y) - 2\pi \hat m_{xy} \frac{y}{\gcd(L_x,L_y)}~,
\\
&\hat \phi_y(\tau+\tfrac12,x,y+\tfrac12) \rightarrow \hat \phi_y(\tau+\tfrac12,x,y+\tfrac12) + 2\pi \hat m_{xy} \frac{x}{\gcd(L_x,L_y)}~,
\\
&\hat n_{xy}(\tau+\tfrac12,x+\tfrac12,y+\tfrac12) \rightarrow \hat n_{xy}(\tau+\tfrac12,x+\tfrac12,y+\tfrac12) - \frac{\hat m_{xy}}{\gcd(L_x,L_y)} \left( L_x \delta_{x,L_x-1} - L_y \delta_{y,L_y-1} \right)~,
\fe
where $\hat m_{xy} = 0,1,\cdots, \text{gcd}(L_x,L_y)-1$. Note that the $\hat m_{xy}=\text{gcd}(L_x,L_y)$ transformation is trivial because it can be undone by an integer gauge transformation with $\hat k_x= -y, \hat k_y= x$. More intuitively, the $\hat m_{xy}=\text{gcd}(L_x,L_y)$ transformation is trivial because $\hat \phi_i$ are compact.
\end{itemize}
The electric and magnetic dipole symmetries do not commute with each other, which leads to a ground state degeneracy of $\gcd(L_x,L_y)$. See the discussion around \eqref{Anoncommute} for more details. In fact, every state in the Hilbert space is $\gcd(L_x,L_y)$-fold degenerate.

There are also time-like electric symmetries:
\begin{itemize}
\item The $U(1)$ time-like electric (dual winding) symmetry acts as
\ie
\mathcal A_\tau \rightarrow \mathcal A_\tau + \frac{c_\tau}{L_\tau}~,
\fe
where $c_\tau \sim c_\tau + 2\pi$ is a circle-valued constant. The charge, which is commonly referred to as the ``gauge charge'', associated with this time-like global symmetry generated by $c_{\tau }$ is a scalar in space, and hence the name scalar charge theory.

\item The $\mathbb Z_{L_x}$ time-like electric (dual winding) dipole symmetry acts as
\ie
&\mathcal A_\tau(\tau+\tfrac12,x,y) \rightarrow \mathcal A_\tau(\tau+\tfrac12,x,y) + 2\pi m_{\tau x} \delta_{\tau,0} \frac{x}{L_x}~,
\\
&n_{\tau xx}(\tau+\tfrac12,x,y) \rightarrow n_{\tau xx}(\tau+\tfrac12,x,y) - m_{\tau x} \delta_{\tau,0} \left( \delta_{x,0} - \delta_{x,L_x-1} \right)~,
\fe
where $m_{\tau x} = 0,\ldots,L_x-1$.

\item The $\mathbb Z_{L_y}$ time-like electric (dual winding) dipole symmetry acts in a similar way with $x$ and $y$ exchanged.
\end{itemize}
As a consequence of the time-like dipole symmetry, the particle described by the defect
\ie\label{2ddipA-defect}
\exp\left( i \sum_{\tau\text{-link: fixed }x,y} \mathcal A_\tau \right)~,
\fe
cannot move, i.e., it is a fracton. The immobility of the fracton in the scalar charge theory is usually attributed to ``dipole moment conservation'' as discussed in \cite{Pretko:2016kxt,Pretko:2016lgv,Pretko:2018jbi}. Here, we give a more precise explanation in terms of the selection rules imposed by the time-like global symmetries.

\subsubsection{Robustness}
Let us impose the $U(1)^3$ electric and $\mathbb Z_{\gcd(L_x,L_y)}$ electric dipole symmetries. Then, the magnetic symmetry and the ground state degeneracy are not robust because the monopole operators $e^{i\hat \phi_i}$ acts nontrivially on the ground states.

Alternatively, we can start with the dual matter theory and impose the dual momentum (magnetic) symmetries. In this case, the dual winding (electric) symmetry is robust because the operators charged under it are extended. It follows that the ground state degeneracy of the matter theory is also robust once we impose the dual momentum symmetries.

\section{Discussion, Conclusion, and Outlook}

In this work, we studied two lattice regularizations of the 2+1d compact Lifshitz theory \cite{Henley1997,Moessner2001,Vishwanath:2004,Fradkin:2004,Ardonne:2003wa,Ghaemi2005,Chen:2009ka,2018PhRvB..98l5105M,Yuan:2019geh,
Lake:2022ico} using the modified Villain formulation of \cite{Sulejmanpasic:2019ytl,Gorantla:2021svj}.  The two models have the same naive continuum Lagrangians given by \eqref{intro:lif-lag}.

Surprisingly, the differences between the two lattice models are significant enough that their low-energy limits are actually distinct and the naive conclusion that both are described by \eqref{intro:lif-lag} is imprecise. In particular, the two models have different global symmetries, anomalies, ground state degeneracies, dualities, etc. 

For example, one of the models, referred to as the Laplacian $\phi$-theory, is self-dual, whereas the other one, referred to as the dipole $\phi$-theory, is dual to a lattice version of a gauge theory known as the vector charge theory \cite{Xu:2006,Bulmash:2018lid,Oh:2021gee,Oh:2022klh}.  The latter contains defects that capture the worldlines of particles that can move only along a line, i.e., lineons. As the name suggests, the dipole $\phi$-theory has a dipole global symmetry which has received a lot of attention in the recent years in the context of fractons. Furthermore, the duality between the dipole $\phi$-theory and the vector charge theory is a rigorous lattice version of the lineon-elasticity duality in the continuum \cite{Manoj:2020bcz}.

We also studied the pure gauge theories in 2+1d associated with the global symmetries of the Laplacian and dipole $\phi$-theories. In the former case, we find the Laplacian gauge theory, which hosts defects that describe immobile particles, i.e., fractons. On the other hand, the latter gives a lattice version of the scalar charge theory \cite{Pretko:2016kxt,Pretko:2016lgv,Bulmash:2018lid,Pretko:2018jbi}, which is also known to host defects that describe fractons. However, the two gauge theories differ in several aspects. For example, while the Laplacian gauge theory has no duality, the scalar charge theory is dual to an elasticity theory of displacements. The latter duality is a rigorous lattice version of the fracton-elasticity duality in the continuum \cite{Pretko:2018qru,Pretko:2019omh,Nguyen:2020yve}.

We emphasize that most of our discussion, including the ``fractonic'' nature of the defects in the gauge theories, is not an artifact of the ``discretization'' of the time direction.  The relation between lattice models with discrete spacetime and lattice models with discrete space and continuous time is well understood.  One way to see that is to consider the Hamiltonian version of these modified Villain models, as in \cite{Yoneda:2022qpj,Cheng:2022sgb,Fazza:2022fss}.

Alternatively, we can take the continuum limit in the time direction as follows.  Consider the scalar charge theory of Section \ref{sec:2ddipA}. Integrating out the scalar fields $\hat \phi^i$ in the action \eqref{2ddipA-modVill-action} imposes the flatness of the integer gauge fields $(n_{\tau ij},n_i)$. We can gauge fix $n_{\tau ij} = 0$ everywhere except at $\tau = \tau_0$, using the integer gauge parameters $k_{ij}$. In other words, after gauge fixing,
\ie
n_{\tau ij}(\tau,x,y) = \delta_{\tau,\tau_0} \bar n_{\tau ij}(x,y)~.
\fe
Flatness then implies that $n_i$'s are independent of $\tau$. Now, we introduce the lattice spacing $a_\tau$ in the time direction and take the limit $a_\tau \rightarrow 0$, while keeping
\ie
\frac{1}{g_0^2} = a_\tau \gamma_0~,\qquad \frac{1}{g'^2_0} = a_\tau \gamma'_0~,\qquad \frac{1}{g^2} = \frac{\gamma}{a_\tau}~,
\fe
fixed. We also scale the field $\mathcal A_\tau$ so that $A_\tau \equiv \frac{1}{a_\tau} \mathcal A_\tau$ is fixed. In this limit, the action \eqref{2ddipA-modVill-action} becomes\footnote{We use $\tau$ to denote the continuum time in this action, but $(x,y)$ still labels the sites on the spatial lattice.}
\ie
S = \oint d\tau~\left[ \frac{1}{2g_0^2} \sum_\text{plaq} E_{xy}^2 + \frac{1}{2g'^2_0} \sum_\text{site} (E_{xx}^2 + E_{yy}^2) + \frac{1}{2g^2} \left( \sum_{x\text{-link}} \mathcal B_x^2 + \sum_{y\text{-link}} \mathcal B_y^2 \right)\right]~,
\fe
where the sums inside the brackets are over the spatial lattice and we defined
\ie
E_{ij} = \partial_\tau \mathcal A_{ij} - \Delta_i \Delta_j A_\tau - 2\pi\delta(\tau - \tau_0) \bar n_{\tau ij}(x,y)~.
\fe
Note that the magnetic fields $\mathcal B_x$ and $\mathcal B_y$ in \eqref{2ddipA-fieldstrengths} are well-defined with continuous $\tau$ because $n_i$'s are independent of $\tau$. The defect \eqref{2ddipA-defect} then becomes
\ie
\exp\left( i\oint d\tau~ A_\tau(\tau,x,y) \right)~.
\fe
It still describes the worldline of a fracton, i.e., an immobile particle, because of the time-like $\mathbb Z_{L_x} \times \mathbb Z_{L_y}$ electric dipole symmetry:
\ie
&A_\tau(\tau,x,y) \rightarrow A_\tau(\tau,x,y) + \delta(\tau-\tau_0) \left( 2\pi m_{\tau x} \frac{x}{L_x} + 2\pi m_{\tau y} \frac{y}{L_y} \right)~,
\\
&\bar n_{\tau xx}(x,y) \rightarrow \bar n_{\tau xx}(x,y) - m_{\tau x} \left( \delta_{x,0} - \delta_{x,L_x-1} \right)~,
\\
&\bar n_{\tau yy}(x,y) \rightarrow \bar n_{\tau yy}(x,y) - m_{\tau y} \left( \delta_{y,0} - \delta_{y,L_y-1} \right)~,
\fe
where $m_{\tau i} = 0,\ldots, L_i -1$.

One can take a similar continuum limit in the time direction for the Laplacian gauge theory of Section \ref{sec:2dlapA} or the vector charge theory of Section \ref{sec:lineonelasticity} without changing any of our conclusions about the mobility of the defects.

In an upcoming paper \cite{Gorantla:2022pii}, we will discuss the $\mathbb Z_N$ version of the $U(1)$ Laplacian gauge theory, and compare it with the 2+1d rank-2 $\mathbb Z_N$ tensor gauge theory \cite{Bulmash:2018lid,Ma:2018nhd,Oh:2021gee,Oh:2022klh,Pace2022}.
We will then consider an  anisotropic generalization of the $\mathbb{Z}_N$ Laplacian model that can be defined on any spatial lattice of the form $ \Gamma\times \mathbb{Z}_{L_z}$, where $\Gamma$ is a general graph.
We will present this lineon model both in terms of a modified Villain lattice action (or more precisely, an integer $BF$ model \cite{Gorantla:2021svj}), and as the low-energy limit of a stabilizer code. The stabilizer code is
\ie\label{stabilizer}
H = -\gamma_1 \sum_{i,z} G(i,z) - \gamma_2 \sum_{i,z} F(i,z+\tfrac12) + \text{c.c.}~,
\fe
where
\ie
&G(i,z) = V_z(i,z+\tfrac12)^\dagger V_z(i,z-\tfrac12) \prod_{j:\langle i, j \rangle\in \Gamma} V(i,z) V(j,z)^\dagger~,
\\
&F(i,z+\tfrac12) = U(i,z+1)^\dagger U(i,z) \prod_{j:\langle i, j \rangle\in \Gamma} U_z(i,z+\tfrac12) U_z(j,z+\tfrac12)^\dagger~.
\fe
where $i,j$ label the sites on  $\Gamma$, and $\langle i,j\rangle \in\Gamma$ means that $i$ and $j$ are connected by an edge.
Here $U(i,z), V(i,z)$ are conjugate $\mathbb{Z}_N$ variables on the sites, i.e., $U(i,z)V(i,z) = e^{2\pi i/N}V(i,z)U(i,z)$.
Similarly, $U_z(i,z+\tfrac12),V_z(i,z+\tfrac12)$ are conjugate variables living on the $z$-links, i.e., $U_z(i,z+\tfrac12)V_z(i,z+\tfrac12) = e^{2\pi i/N}V_z(i,z+\tfrac12)U_z(i,z+\tfrac12)$.
This anisotropic $\mathbb{Z}_N$ lattice model is gapped, robust, and has lineons.
Its GSD on a general graph is given by $|\Jac(\Gamma, N)|^2$, where $\Jac(\Gamma,N)$ is a mod $N$ reduction of the Jacobian group $\Jac(\Gamma)$ of  $\Gamma$.
A special case of $\Gamma$ corresponds to a square lattice with sizes $L_x,L_y$.  When $N$ is prime, the logarithm  of the GSD is given by the dimension of a quotient ring (as a vector space):
\ie\label{GSDtorus}
\log_N \text{GSD} = 2\dim_{\mathbb{Z}_N} { \mathbb{Z}_N[X,Y] \over \left( \, Y(X-1)^2 +X(Y-1)^2 , X^{L_x}-1, Y^{L_y}-1 \, \right) } \,.
\fe
In this case, as in the celebrated Haah's code \cite{Haah:2011drr}, the logarithm of the GSD depends in a complicated, non-monotonic way on $L_x,L_y$ and grows at most linearly in the system size. Furthermore, there are sequences of $L_x,L_y$ going to infinity such that the GSD stays finite.

\textit{Notes added:} As we were finishing this paper, \cite{Ebisu:2022nln} appeared on the arXiv.  It overlaps with some of the  findings in our upcoming paper \cite{Gorantla:2022pii}.
In particular, the stabilizer code \eqref{stabilizer} and the GSD formula on a general graph appear in \cite{Ebisu:2022nln}. In \cite{Gorantla:2022pii}, we will investigate this model also on a cubic lattice and derive \eqref{GSDtorus}.

\section*{Acknowledgements}
We are grateful to L.\ Radzihovsky, T.\ Senthil, and A.\ Vishwanath for useful discussions. PG was supported by the Physics Department of Princeton University. HTL is supported in part by a Croucher fellowship from the Croucher Foundation, the Packard Foundation and the Center for Theoretical Physics at MIT. The work of NS was supported in part by DOE grant DE$-$SC0009988 and by the Simons Collaboration on Ultra-Quantum Matter, which is a grant from the Simons Foundation (651440, NS).  The work of SHS was supported in part by NSF grant PHY-2210182. The authors of this paper were ordered alphabetically.

\appendix

\section{More on global symmetries of 2+1d dipole theories}\label{app:2ddip}
In this appendix, we will analyze the global symmetries of the modified Villain versions of the 2+1d dipole $\phi$-theory of Section \ref{sec:2ddipphi} and the 2+1d scalar charge theory of Section \ref{sec:2ddipA} in more detail. We will discuss the charges/symmetry operators, charged operators/defects. We will also discuss the ground state degeneracy as a consequence of space-like symmetries and restricted mobility of defects as a consequence of time-like symmetries.

\subsection{Global symmetry of 2+1d dipole $\phi$-theory}\label{app:2ddipphi}

The global symmetries of the modified Villain model \eqref{2ddipphi-modVill-action} are listed below:
\begin{itemize}
\item
The $U(1)$ momentum (dual magnetic) symmetry acts as $\phi \rightarrow \phi + c$, where $c\sim c+2\pi$. The Noether currents are
\ie\label{2ddipphi-modVill-momcur}
J_\tau = i\beta_0 (\Delta_\tau \phi - 2\pi n_\tau)~,\qquad J_{xy} = i\beta (\Delta_x \Delta_y \phi - 2\pi n_{xy})~,\qquad J_{ii} = i\beta' (\Delta_i^2 \phi - 2\pi n_{ii})~,
\fe
and they satisfy
\ie
\Delta_\tau J_\tau = \Delta_i \Delta_j J_{ij}~,
\fe
which follows from the equation of motion of $\phi$. The charge is
\ie
Q = \sum_{\tau\text{-link: fixed }\tau} J_\tau~,
\fe
and the charged operator is $e^{i \phi}$.

\item
The $\mathbb Z_{L_x}$ momentum (dual magnetic) dipole symmetry acts as
\ie\label{2ddipphi-modVill-momdip}
&\phi \rightarrow \phi + 2\pi m_{xx} \frac{x}{L_x}~,
\\
&n_{xx} \rightarrow n_{xx} + m_{xx} \left( \delta_{x,0} - \delta_{x,L_x-1} \right)~,
\fe
where $m_{xx}=0,1,\ldots, L_x-1$. The symmetry operator is
\ie\label{2ddipphi-modVill-momdip-symop}
U_{m_{xx}}^{(x)} &= \exp\left( \frac{2\pi i m_{xx}}{L_x} \sum_{\tau\text{-link: fixed }\tau} x J_\tau - i m_{xx} \sum_{\tau\text{-link: fixed }\tau}\left[ \hat{\mathcal A}_{yy}(x = 0) - \hat{\mathcal A}_{yy}(x = L_x-1) \right] \right)
\\
&=\exp\left( -\frac{2\pi i m_{xx}}{L_x} \sum_{\tau\text{-link: fixed }\tau} x \hat n \right)~.
\fe
The charged operators are $e^{i\phi}$, and the dipole operator $e^{i \Delta_x \phi}$.\footnote{These operators and their correlation functions have been discussed extensively in the context of spontaneous symmetry breaking of the ordinary and dipole symmetries in \cite{Stahl:2021sgi,Lake:2022ico,Kapustin:2022fzp}.}

\item
The $\mathbb Z_{L_y}$ momentum (dual magnetic) dipole symmetry acts in a similar way with $x$ and $y$ exchanged.

\item
There is a $U(1)^3$ winding (dual electric) symmetry that shifts\footnote{Naively, it might appear that $\hat{\mathcal A}_{xy}$ can be shifted by $\hat c^x(x) +\hat c^y(y)$, but in fact these shifts can be gauged away except for the zero mode.}
\ie
\hat{\mathcal A}_{ii} \rightarrow \hat{\mathcal A}_{ii} + \frac{\hat c_{ii}}{L_i}~,\qquad \hat{\mathcal A}_{xy} \rightarrow \hat{\mathcal A}_{xy} + \frac{\hat c_{xy}}{\lcm(L_x,L_y)}~,
\fe
where $\hat c_{ij}\sim \hat c_{ij} + 2\pi$. The Noether currents are
\ie
\hat J_{\tau ij} = \frac{1}{2\pi} (\Delta_i \Delta_j \phi - 2\pi n_{ij})~,\qquad \hat J = \frac{1}{2\pi} (\Delta_\tau \phi - 2\pi n_\tau)~,
\fe
and they satisfy
\ie
\Delta_\tau \hat J_{\tau ij} = \Delta_i \Delta_j \hat J~, \qquad \Delta_x \hat J_{\tau xy} = \Delta_y \hat J_{\tau xx}~,\qquad \Delta_y \hat J_{\tau xy} = \Delta_x \hat J_{\tau yy}~,
\fe
which follow from the equations of motion of $\hat{\mathcal A}_{ij}$ and $\hat{\mathcal A}_{\tau i}$. The first equation is the conservation equation, and the last two equations are difference conditions (Gauss laws). The charges are
\ie
&\hat Q_{xx}(y) = \sum_{\text{site: fixed }\tau,y} \hat J_{\tau xx} = -\sum_{\text{site: fixed }\tau,y} n_{xx}~,
\\
&\hat Q_{yy}(x) = \sum_{\text{site: fixed }\tau,x} \hat J_{\tau yy} = -\sum_{\text{site: fixed }\tau,x} n_{yy}~,
\\
&\hat Q_{xy}^x(x) = \sum_{xy\text{-plaq: fixed }\tau,x} \hat J_{\tau xy} = -\sum_{xy\text{-plaq: fixed }\tau,x} n_{xy}~,
\\
&\hat Q_{xy}^y(y) = \sum_{xy\text{-plaq: fixed }\tau,y} \hat J_{\tau xy} = -\sum_{xy\text{-plaq: fixed }\tau,y} n_{xy}~.
\fe
The difference conditions (Gauss laws) imply that the four charges are independent of their arguments. Moreover, the last two charges are related as
\ie
-\sum_{xy\text{-plaq: fixed }\tau} n_{xy} = L_x \hat Q_{xy}^x = L_y \hat Q_{xy}^y = \lcm(L_x,L_y) \hat Q_{xy}~,
\fe
where $\hat Q_{xy}$ is an integer. So there are only three independent $U(1)$ charges, $\hat Q_{ij}$.

The charged operators are
\ie\label{2ddipphi-modVill-U1-wind-chargedop}
&\hat W_{yy}(\tau+\tfrac12,x) = \exp\left( i\sum_{\tau\text{-link: fixed }\tau,x} \hat{\mathcal A}_{yy} \right)~,
\\
&\hat W_{xx}(\tau+\tfrac12,y) = \exp\left( i\sum_{\tau\text{-link: fixed }\tau,y} \hat{\mathcal A}_{xx} \right)~,
\\
&\hat W_{xy}(\tau+\tfrac12) = \exp\left( i \sum_{\text{cube: fixed }\tau} \hat{\mathcal A}_{xy} \right)~,
\fe
respectively. Actually, the third operator in \eqref{2ddipphi-modVill-U1-wind-chargedop} is not minimally charged under the $U(1)$ winding symmetry generated by $\hat c_{xy}$. Instead, the minimally charged operator is the ``diagonal'' operator
\ie\label{2ddipphi-modVill-diagop}
&\exp\Bigg( i \sum_{s=0}^{\lcm(L_x,L_y)-1} \left[ \hat{\mathcal A}_{xy}(\tau+\tfrac12,x+s+\tfrac12,y+s+\tfrac12) \right.
\\
&\qquad \qquad +\left. \hat{\mathcal A}_{xx}(\tau+\tfrac12,x+s,y+s) + \hat{\mathcal A}_{yy}(\tau+\tfrac12,x+s,y+s) \right]\Bigg)~,
\fe
which, however, is also charged under the other $U(1)$ winding symmetries generated by $\hat c_{ii}$. A different minimally charged operator, which is not charged under the other $U(1)$ winding symmetries is
\ie
\exp\left[ \frac{i}{\gcd(L_x,L_y)}\sum_{x,y} \left(\hat{\mathcal A}_{xy}(\tau+\tfrac12,x+\tfrac12,y+\tfrac12) - 2\pi x y~ \hat n(\tau+\tfrac12,x,y) \right) \right]~.
\fe

\item
There is a $\mathbb Z_{L_x}$ winding (dual electric) dipole symmetry that shifts
\ie
&\hat{\mathcal A}_{yy} \rightarrow \hat{\mathcal A}_{yy} + 2\pi \hat m_{xx} \frac{x}{L_x} \delta_{y,0}~,
\\
&\hat n \rightarrow \hat n - \hat m_{xx} \delta_{y,0} (\delta_{x,0} - \delta_{x,L_x-1})~.
\fe
where $\hat m_{xx} = 0,\ldots,L_x-1$. The symmetry operator is
\ie
\hat U^{(x)}_{\hat m_{xx}} = \exp \left( -\frac{2\pi i \hat m_{xx}}{L_x} \sum_{\text{site: fixed }\tau} x n_{xx}(y = 0) \right)~.
\fe
The charged operators are $\hat W_{yy}(\tau+\tfrac12,x)$, and $\hat W_{yy}(\tau+\tfrac12,x+1) \hat W_{yy}(\tau+\tfrac12,x)^{-1}$. The diagonal operator \eqref{2ddipphi-modVill-diagop} is also charged under this symmetry.

\item
There is also a $\mathbb Z_{L_y}$ winding (dual electric) dipole symmetry, which acts in a similar way with $x$ and $y$ exchanged.

\end{itemize}
The momentum (dual magnetic) and winding (dual electric) dipole symmetries do not commute:
\ie\label{phinoncommute}
U_{m_{xx}}^{(x)} \hat U^{(x)}_{\hat m_{xx}} = e^{-\frac{2\pi i}{L_x} m_{xx} \hat m_{xx}}\hat U^{(x)}_{\hat m_{xx}} U_{m_{xx}}^{(x)}~,
\fe
and similarly in the $y$-direction. This signals a mixed 't Hooft anomaly between them and leads to a large $L_x L_y$-fold ground state degeneracy.

There are also time-like winding (dual electric) symmetries that act on the defects of $(\hat{\mathcal A}_{\tau i},\hat{\mathcal A}_{ij})$:
\begin{itemize}

\item
The $U(1)^2$ time-like winding (dual electric) symmetry acts as
\ie
\hat{\mathcal A}_{\tau i} \rightarrow \hat{\mathcal A}_{\tau i} + \frac{\hat c_{\tau i}}{L_\tau}~,
\fe
where $\hat c_{\tau i} \sim \hat c_{\tau i} + 2\pi$ is circle-valued. The charged defects are
\ie
\hat W_{\tau x}(x+\tfrac12,y) = \exp\left( i \sum_{x\text{-link: fixed }x,y} \hat{\mathcal A}_{\tau x} \right)~,
\fe
and similarly $\hat W_{\tau y}(x,y+\tfrac12)$. The defect $\hat W_{\tau x}(x+\tfrac12,y)$ describes the world-line of a particle on the $x$-link $(x+\tfrac12,y)$. Since it can move in the $x$-direction via $\hat{\mathcal A}_{xx}$, we call it the $x$-lineon. Similarly, the defect $\hat W_{\tau y}(x,y+\tfrac12)$ describes the world-line of a $y$-lineon.

\item
The $\mathbb Z_{\gcd(L_x,L_y)}$ time-like winding (dual electric) dipole symmetry acts as
\ie
&\hat{\mathcal A}_{\tau x}(\tau,x+\tfrac12,y) \rightarrow \hat{\mathcal A}_{\tau x}(\tau,x+\tfrac12,y) - 2\pi \hat m_{\tau xy} \delta_{\tau,0} \frac{y}{\gcd(L_x,L_y)}~,
\\
&\hat{\mathcal A}_{\tau y}(\tau,x,y+\tfrac12) \rightarrow \hat{\mathcal A}_{\tau y}(\tau,x,y+\tfrac12) + 2\pi \hat m_{\tau xy} \delta_{\tau,0} \frac{x}{\gcd(L_x,L_y)}~,
\\
&\hat n_{\tau xy}(\tau,x+\tfrac12,y+\tfrac12) \rightarrow \hat n_{\tau xy}(\tau,x+\tfrac12,y+\tfrac12) + \frac{\hat m_{\tau xy} \delta_{\tau,0}}{\gcd(L_x,L_y)} \left( L_x \delta_{x,L_x-1} - L_y \delta_{y,L_y-1} \right)~,
\fe
where $\hat m_{\tau xy} = 0,\ldots,\gcd(L_x,L_y)-1$.

\end{itemize}
These time-like symmetries imply that the defects $\hat W_{\tau x}(x+\tfrac12,y)$ and $\hat W_{\tau x}(x'+\tfrac12,y')$ have the same time-like charges if and only if
\ie
(x',y') = (x + s_x, y + \gcd(L_x,L_y) s_y)~,
\fe
where $s_x,s_y$ are integers. In other words, an $x$-lineon can move anywhere in the $x$-direction, but it can hop only by $\gcd(L_x,L_y)$ sites in the $y$-direction. On the other hand, a dipole of $x$-lineons separated in the $y$-direction, described by the defect $\hat W_{\tau x}(x+\tfrac12,y') \hat W_{\tau x}(x+\tfrac12,y)^{-1}$, is fully mobile. Similar mobility restrictions apply to the $y$-lineons. Interestingly, a pair of $x$- and $y$-lineons on the links $(x+\tfrac12,y)$ and $(x,y+\tfrac12)$ respectively can move ``diagonally'' together to $(x+s+\tfrac12,y+s)$ and $(x+s,y+s+\tfrac12)$ respectively, for any $s\in\mathbb Z$.

\subsection{Global symmetry of 2+1d scalar charge theory}\label{app:2ddipA}
The global symmetries of the modified Villain model \eqref{2ddipA-modVill-action} are listed below:
\begin{itemize}
\item
The $U(1)^3$ electric (dual winding) symmetry acts as
\ie
\mathcal A_{ii} \rightarrow \mathcal A_{ii} + \frac{c_{ii}}{L_i}~,\qquad \mathcal A_{xy} \rightarrow \mathcal A_{xy} + \frac{c_{xy}}{\gcd(L_x,L_y)}~,
\fe
where $c_{ij} \sim c_{ij} + 2\pi$ are circle-valued constants. The Noether currents are
\ie
&J_{\tau ii} = -i\gamma_0' \mathcal E_{ii}~,\qquad J_{\tau xy} = i\gamma_0 \mathcal E_{xy}~,
\\
&J_x = i\gamma \mathcal B_x~,\qquad J_y = i\gamma \mathcal B_y~,
\fe
and they satisfy
\ie
&\Delta_\tau J_{\tau xx} = \Delta_y J_y~,\qquad && \Delta_\tau J_{\tau xy} = \Delta_y J_x + \Delta_x J_y~,
\\
&\Delta_\tau J_{\tau yy} = \Delta_x J_x~,\qquad && \Delta_x \Delta_y J_{\tau xy} - \Delta_y^2 J_{\tau yy} - \Delta_x^2 J_{\tau xx} = 0~.
\fe
They follow from the equations of motion of $\mathcal A_{ij}$ and $\mathcal A_\tau$. The first three equations are the conservation equations, and the last equation is the difference condition (Gauss law). The charges are
\ie
Q_{ii}(x^i) = \sum_{\tau\text{-link: fixed }\tau, x^i} J_{\tau ii}~,\qquad\qquad Q_{xy} = \sum_{\text{cube: fixed }\tau} J_{\tau xy}~.
\fe
The difference condition (Gauss law) implies that $Q_{ii}$ is independent of $x^i$, while $Q_{xy}$ is a multiple of $\gcd(L_x,L_y)$.\footnote{This can be seen easily in the dual frame \eqref{2ddipA-modVill-dual-action}, where the charge is $Q_{xy} = -\sum_{\text{cube: fixed }\tau} \hat n_{xy}$.} The charged operators are
\ie\label{2ddipA-modVill-U1-elec-chargedop}
&W_{xx}(\tau,y) = \exp\left( i \sum_{\text{site: fixed }\tau,y} \mathcal A_{xx} \right)~,\qquad W_{yy}(\tau,x) = \exp\left( i \sum_{\text{site: fixed }\tau,x} \mathcal A_{yy} \right)~,
\\
&W^x_{xy}(\tau,x+\tfrac12) = \exp\left( i \sum_{xy\text{-plaq: fixed }\tau,x} \mathcal A_{xy} \right)~,\qquad W^y_{xy}(\tau,y+\tfrac12) = \exp\left( i \sum_{xy\text{-plaq: fixed }\tau,y} \mathcal A_{xy} \right)~,
\fe
respectively. Actually, the two operators in the second line of \eqref{2ddipA-modVill-U1-elec-chargedop} are not minimally charged under the $U(1)$ electric symmetry generated by $c_{xy}$. Instead, the minimally charged operator is
\ie
&\exp\Bigg[ \frac{i}{\lcm(L_x,L_y)} \sum_{x,y} \bigg(\mathcal A_{xy}(\tau,x+\tfrac12,y+\tfrac12)
\\
&\qquad \qquad -  \frac{2\pi \tilde L_y L_y x}{\gcd(L_x,L_y)} n_y(\tau,x,y+\tfrac12) - \frac{2\pi \tilde L_x L_x y}{\gcd(L_x,L_y)} n_x(\tau,x+\tfrac12,y) \bigg) \Bigg]~,
\fe
where $\tilde L_i$ are integer solutions of the equation $\tilde L_x L_x + \tilde L_y L_y = \gcd(L_x,L_y)$. Another minimally charged operator is
\ie
W^x_{xy}(\tau,x+\tfrac12)^{\tilde L_y} W^y_{xy}(\tau,y+\tfrac12)^{\tilde L_x}~.
\fe

\item
The $\mathbb Z_{\gcd(L_x,L_y)}$ electric (dual winding) dipole symmetry acts as
\ie
&\mathcal A_{xy}(\tau,x+\tfrac12,y+\tfrac12) \rightarrow \mathcal A_{xy}(\tau,x+\tfrac12,y+\tfrac12) + \frac{2\pi m_{xy}}{\gcd(L_x,L_y)} \left(\tilde L_y \delta_{x,L_x-1} - \tilde L_x \delta_{y,L_y-1}\right)~,
\\
&\mathcal A_{xx}(\tau,x,y) \rightarrow \mathcal A_{xx}(\tau,x,y) - \frac{2\pi m_{xy}}{\gcd(L_x,L_y)} \tilde L_y y (\delta_{x,0} - \delta_{x,L_x-1})~,
\\
&\mathcal A_{yy}(\tau,x,y) \rightarrow \mathcal A_{yy}(\tau,x,y) + \frac{2\pi m_{xy}}{\gcd(L_x,L_y)} \tilde L_x x (\delta_{y,0} - \delta_{y,L_y-1})~,
\\
&n_x(\tau,x+\tfrac12,y) \rightarrow n_x(\tau,x+\tfrac12,y) - \frac{m_{xy}}{\gcd(L_x,L_y)} \tilde L_x L_x (\delta_{y,0} - \delta_{y,L_y-1}) \delta_{x,L_x-1}~,
\\
&n_y(\tau,x,y+\tfrac12) \rightarrow n_y(\tau,x,y+\tfrac12) + \frac{m_{xy}}{\gcd(L_x,L_y)} \tilde L_y L_y (\delta_{x,0} - \delta_{x,L_x-1}) \delta_{y,L_y-1}~,
\fe
where $m_{xy} = 0,\ldots,\gcd(L_x,L_y)-1$. The symmetry operator is
\ie
U_{m_{xy}} &= \exp\Bigg[\frac{2\pi i m_{xy}}{\gcd(L_x,L_y)} \bigg( \tilde L_x \sum_{x
\atop y=L_y-1} [\hat n_{xy}(\tau+\tfrac12,x+\tfrac12,y+\tfrac12) + x \Delta_y \hat n_{xx}(\tau+\tfrac12,x,y+\tfrac12)]
\\
&\qquad \quad - \tilde L_y \sum_{y\atop x=L_x-1} [\hat n_{xy}(\tau+\tfrac12,x+\tfrac12,y+\tfrac12) + y \Delta_x \hat n_{yy}(\tau+\tfrac12,x+\tfrac12,y)] \bigg) \Bigg]~.
\fe
The charged operators are $W^i_{xy}(\tau,x^i+\tfrac12)$. The minimally charged operator is
\ie
W^y_{xy}(\tau,y+\tfrac12)^{\frac{L_y}{\gcd(L_x,L_y)}} W^x_{xy}(\tau,x+\tfrac12)^{-\frac{L_x}{\gcd(L_x,L_y)}}~,
\fe
or
\ie
\prod_{s_y = 1}^{\frac{L_y}{\gcd(L_x,L_y)}} W^y_{xy}(\tau,y +\gcd(L_x,L_y)s_y+\tfrac12) \prod_{s_x = 1}^{\frac{L_x}{\gcd(L_x,L_y)}} W^x_{xy}(\tau,x +\gcd(L_x,L_y)s_x+\tfrac12)^{-1}~.
\fe

\item
The $U(1)^2$ magnetic (dual momentum) symmetry acts as $\hat \phi_i \rightarrow \hat \phi_i + \hat c_i$, where $\hat c_i \sim \hat c_i + 2\pi$ are circle-valued constants. The Noether currents are
\ie
&\hat J_{\tau x} = \frac{1}{2\pi} \mathcal B_x~,\qquad \hat J_{\tau y} = \frac{1}{2\pi} \mathcal B_y~,
\\
&\hat J_{xx} = -\frac{1}{2\pi} \mathcal E_{yy}~,\qquad \hat J_{yy} = -\frac{1}{2\pi} \mathcal E_{xx}~,\qquad \hat J_{xy} = \frac{1}{2\pi} \mathcal E_{xy}~,
\fe
and they satisfy
\ie
&\Delta_\tau \hat J_{\tau x} + \Delta_x \hat J_{xx} + \Delta_y \hat J_{xy} = 0~,
\\
&\Delta_\tau \hat J_{\tau y} + \Delta_y \hat J_{yy} + \Delta_x \hat J_{xy} = 0~,
\fe
which follow from the equations of motion of $\hat \phi_i$. The charges are
\ie
Q_i = \sum_{i\text{-link: fixed }\tau} \hat J_{\tau i}~,
\fe
and the charged operators are $e^{i \hat \phi_i}$.

\item
The $\mathbb Z_{\gcd(L_x,L_y)}$ magnetic (dual momentum) dipole symmetry acts as
\ie
&\hat \phi_x(\tau+\tfrac12,x+\tfrac12,y) \rightarrow \hat \phi_x(\tau+\tfrac12,x+\tfrac12,y) - 2\pi \hat m_{xy} \frac{y}{\gcd(L_x,L_y)}~,
\\
&\hat \phi_y(\tau+\tfrac12,x,y+\tfrac12) \rightarrow \hat \phi_y(\tau+\tfrac12,x,y+\tfrac12) + 2\pi \hat m_{xy} \frac{x}{\gcd(L_x,L_y)}~,
\\
&\hat n_{xy}(\tau+\tfrac12,x+\tfrac12,y+\tfrac12) \rightarrow \hat n_{xy}(\tau+\tfrac12,x+\tfrac12,y+\tfrac12) - \frac{\hat m_{xy}}{\gcd(L_x,L_y)} \left( L_x \delta_{x,L_x-1} - L_y \delta_{y,L_y-1} \right)~,
\fe
where $\hat m_{xy} = 0,\ldots,\gcd(L_x,L_y)-1$. The symmetry operator is
\ie
\hat U_{\hat m_{xy}} = \exp\left( -\frac{2\pi i \hat m_{xy}}{\gcd(L_x,L_y)} \sum_{x,y} \left[ x n_y(\tau,x,y+\tfrac12) - y n_x(\tau,x+\tfrac12,y) \right] \right)~.
\fe
The charged operators are $e^{i\hat \phi_i}$, and the dipole operators $e^{i\Delta_x \hat \phi_y}$ and $e^{i\Delta_y \hat \phi_x}$.

\end{itemize}
The electric (dual winding) and magnetic (dual momentum) dipole symmetries do not commute:
\ie\label{Anoncommute}
U_{m_{xy}} \hat U_{\hat m_{xy}} = e^{-\frac{2\pi i }{\gcd(L_x,L_y)} m_{xy} \hat m_{xy}} \hat U_{\hat m_{xy}} U_{m_{xy}}~.
\fe
This signals a mixed 't Hooft anomaly between them and leads to a ground state degeneracy of $\gcd(L_x,L_y)$.

There are also time-like electric (dual winding) symmetries that act on the defects of $(\mathcal A_\tau,\hat{\mathcal A}_{ij})$:
\begin{itemize}
\item
The $U(1)$ time-like electric (dual winding) symmetry acts as
\ie
\mathcal A_\tau \rightarrow \mathcal A_\tau + \frac{c_\tau}{L_\tau}~,
\fe
where $c_\tau \sim c_\tau + 2\pi$ is a circle-valued constant. The charged defect is
\ie
W_\tau(x,y) = \exp\left( i \sum_{\tau\text{-link: fixed }x,y} \mathcal A_\tau \right)~.
\fe
It describes the world-line of a particle at $(x,y)$.

\item
The $\mathbb Z_{L_x}$ time-like electric (dual winding) dipole symmetry acts as
\ie
&\mathcal A_\tau(\tau+\tfrac12,x,y) \rightarrow \mathcal A_\tau(\tau+\tfrac12,x,y) + 2\pi m_{\tau x} \delta_{\tau,0} \frac{x}{L_x}~,
\\
&n_{\tau xx}(\tau+\tfrac12,x,y) \rightarrow n_{\tau xx}(\tau+\tfrac12,x,y) - m_{\tau x} \delta_{\tau,0} \left( \delta_{x,0} - \delta_{x,L_x-1} \right)~,
\fe
where $m_{\tau x} = 0,\ldots,L_x-1$.

\item
The $\mathbb Z_{L_y}$ time-like electric (dual winding) dipole symmetry acts in a similar way with $x$ and $y$ exchanged.

\end{itemize}
These time-like symmetries imply that the defects $W_\tau(x,y)$ and $W_\tau(x',y')$ have the same time-like charges if and only if $(x',y') = (x,y)$. In other words, the particle described by the defect $W_\tau$ cannot move, i.e., it is a fracton. On the other hand, any dipole of fractons, described by the defect $W_\tau(x',y') W_\tau(x,y)^{-1}$, is fully mobile.

\section{Polynomial representation of functions on square lattice}\label{app:poly}

In this appendix, we develop a polynomial representation of functions on the infinite square lattice $\mathbb Z^2$, and use it to show the following:
\begin{enumerate}
\item The naturalness of the action \eqref{2dlapphi-modVill-action} with respect to the global symmetry of the 2+1d Laplacian $\phi$-theory of Section \ref{sec:2dlapphi}. More precisely, we show that the local operator $\prod_{i=1}^n e^{iq_i\phi(0,x_i,y_i)}$ is invariant under the momentum symmetry if and only if it can be written as $\prod_{j=1}^m e^{ir_j\Delta_L \phi(0,x_j,y_j)}$, where $q_i,r_j\in\mathbb Z$ and $\Delta_L:= \Delta_x^2+\Delta_y^2$ is the discrete Laplacian operator. Of course the winding operator $e^{i\tilde \phi}$ is invariant under the momentum symmetry, and it is relevant because it acts nontrivially on the ground states, so the action \eqref{2dlapphi-modVill-action} is not natural unless we impose the winding symmetry too.

\item The immobility of a finite set of defects with arbitrary charges, unless they can be ``locally annihilated,'' in the 2+1d $U(1)$ Laplacian gauge theory of Section \ref{sec:2dlapA}.
\end{enumerate}
The polynomial representation was originally developed in the context of translationally invariant Pauli stabilizer codes \cite{Haah_2013}.

On an infinite square lattice $\mathbb Z^2$, any function $f$ can be associated with a formal \emph{Laurent power series} in the variables $X,Y$:
\ie
\hat f(X,Y) = \sum_{(x,y)\in\mathbb Z^2} f(x,y) X^{-x} Y^{-y}~.
\fe
We can think of $X=e^{ip_x}$ and $Y=e^{ip_y}$ as phases with $p_x$ and $p_y$ momenta conjugate to $x$ and $y$.  Then, this definition of $\hat f(X,Y)$ is the discrete Fourier transform of $f(x,y)$.  Related to that,
$X$ and $Y$ are generators of lattice translations in the $x$ and $y$ directions:
\ie
X \hat f(X,Y) = \sum_{(x,y)\in\mathbb Z^2} f(x+1,y) X^{-x} Y^{-y}~,
\fe
and similarly for $Y$. Then, the difference operator $\Delta_x$ is associated with $X-1$ because
\ie
(X-1) \hat f(X,Y) = \sum_{(x,y)\in\mathbb Z^2} \Delta_x f(x+\tfrac12,y) X^{-x} Y^{-y}~.
\fe
Recall that $\Delta_x f(x+\tfrac12,y) = f(x+1,y)-f(x,y)$.

More generally, any \emph{local} difference operator is associated with a \emph{Laurent polynomial} $s(X,Y)$ with integer coefficients, i.e., an element of $\mathbb Z[X,X^{-1},Y,Y^{-1}]$ satisfying $s(1,1)=0$.  (Here, $\mathbb Z[X,Y,\ldots]$ is the set of polynomials in $X,Y,\ldots$ with integer coefficients, and therefore $\mathbb Z[X,X^{-1},Y,Y^{-1},\ldots]$ is the set of Laurent polynomials in $X,Y,\ldots$ with integer coefficients.)

For example, the discrete Laplacian operator $\Delta_L := \Delta_x^2 + \Delta_y^2$ corresponds to the Laurent polynomial
\ie\label{polyp}
p(X,Y)=(X - 2 + X^{-1}) + (Y - 2 + Y^{-1})~.
\fe
We can equivalently work with
\ie\label{polyptilde}
\tilde p(X,Y) = XYp(X,Y) = Y(X-1)^2 + X(Y-1)^2~,
\fe
which is simply a translated version of $\Delta_L$. Note that $\tilde p(X,Y) \in \mathbb Z[X,Y]$, i.e., $\tilde p(X,Y)$ is a polynomial, whereas $p(X,Y)$ is a Laurent polynomial. Indeed, we can always translate a difference operator so that the associated Laurent polynomial is a polynomial.\footnote{In the continuum, a differential operator in space becomes in momentum space a multiplication by a polynomial in the momenta.  On the lattice, we follow the interpretation of $X$ and $Y$ as lattice translation generators, i.e., $X=e^{ip_x}$, $Y=e^{ip_y}$, and then a difference operator becomes a polynomial in $X$ and $Y$.}

Let us define a \emph{lexicographic monomial order}, $X\succ Y$, on $\mathbb Z[X,Y]$. This means we can compare any two monomials as follows: $X^m Y^n \succ X^k Y^l$ if $m>k$, or $m=k$ and $n>l$. Clearly, this is a total order on all monomials in $X,Y$. Given a nonzero polynomial, its \emph{leading term} is the term with the largest monomial among all its terms. The corresponding coefficient and monomial are called \emph{leading coefficient} and \emph{leading monomial} respectively. If the leading coefficient is $\pm1$, the polynomial is said to be \emph{monic}.

We say a polynomial $a(X,Y)$ is \emph{reducible} by another polynomial $b(X,Y)$ if some term of $a(X,Y)$ is a multiple of the leading term of $b(X,Y)$. Furthermore, if $b(X,Y)$ is monic, then $a(X,Y)$ can be written uniquely as
\ie\label{rdefini}
a(X,Y) = c(X,Y) b(X,Y) + d(X,Y)~,
\fe
where $c(X,Y)$ is the \emph{quotient} and $d(X,Y)$ is the \emph{remainder}, which are uniquely determined by demanding that $d(X,Y)$ is not reducible by $b(X,Y)$. This operation is known as \emph{multivariate division} with respect to a given monomial order.

\subsection{Naturalness of 2+1d Laplacian $\phi$-theory}\label{app:polyphi}

In this appendix, we show that the action \eqref{2dlapphi-modVill-action} is natural with respect to the global symmetry of the 2+1d Laplacian $\phi$-theory.

Usually, the notion of naturalness assumes that a certain global symmetry is imposed on the system and then all the relevant operators in the action respect this symmetry \cite{tHooft:1979rat}. (See \cite{paper1} for a more recent discussion comparing the notions of naturalness and robustness.)  For this to make sense, we need some scaling property, which determines which terms in the action should be viewed as relevant, and which terms should be viewed as irrelevant.  In the lattice system, without a continuum limit, there is no such obvious scaling.  Instead, we show that every term that respects the symmetry can be expressed in terms of lattice derivatives acting on other terms that are already present in the action.  More precisely, we will show that every term invariant under the momentum symmetry can be expressed in terms of gauge invariant functions of $\Delta_L\phi$ and lattice derivatives of them.  Then, we will exclude more terms using the winding symmetry. See more details below.

In the continuum, the conclusion of this appendix is the following trivial statement.  A differential operator $\mathcal D $ that annihilates every real harmonic function on $\mathbb R^2$, $f(x,y)$ includes the Laplacian as a factor.  To see that, use holomorphic coordinates and write $f=g(z)+\bar g(\bar z)$.  Then, $\mathcal D f=0$ means that $\mathcal D $ must include a factor of $\partial_z$ and using the reality, it should also have a factor of $\partial_{\bar z}$.  Therefore, $\mathcal D =\mathcal{D}' \partial_z\partial_{\bar z}$ with a differential operator $\mathcal {D}' $.

The momentum symmetry of the 2+1d Laplacian $\phi$-theory on the square lattice includes shifts by real-valued discrete harmonic functions $f(x,y)$ on $\mathbb Z^2$ (see Section \ref{sec:2dlapphi-sym}). We would like to find other terms invariant under this symmetry.  We look for terms depending on $D\phi$ with some difference operator $D$.  Let $\mathcal H(\mathbb Z^2,\mathbb R)$ be the set of all real-valued discrete harmonic functions. By definition, any $f\in\mathcal H(\mathbb Z^2,\mathbb R)$ satisfies
\ie
\Delta_L f(x,y) = 0 \iff p(X,Y) \hat f(X,Y) = 0~.
\fe
We would like to find the condition that the difference operator $D$ should satisfy such that $Df(x,y)=0$ for all $f\in\mathcal H(\mathbb Z^2,\mathbb R)$.\footnote{We should impose $Df=0$ rather than the weaker condition $Df(x,y) = 0 \mod 2\pi$ because it should be true for $c f(x,y)$ for any $c\in\mathbb R$.\label{cannotimp}}

One trivial possibility is $D = D'\circ \Delta_L$ because $(D' \circ \Delta_L)f(x,y) = 0$ for any operator $D'$. This means, we could add to the action \eqref{2dlapphi-modVill-action} a term of the form
\ie
-\sum_\text{site} \cos[(D'\circ \Delta_L) \phi]~,
\fe
and preserve the global symmetry. This is considered a higher-order term than $\Delta_L$ and is always compatible with the momentum global symmetry.

A more interesting possibility would be a $D$ that cannot be written as $D' \circ \Delta_L$, and yet $Df(x,y)= 0$ for all $f\in\mathcal H(\mathbb Z^2,\mathbb R)$. Below, we will show that this is impossible. Equivalently, we show that any $D$ that satisfies $Df(x,y) = 0$ for all $f\in\mathcal H(\mathbb Z^2,\mathbb R)$ is of higher order than $\Delta_L$. This implies that the action \eqref{2dlapphi-modVill-action} is natural with respect to the global momentum symmetry of the 2+1d Laplacian $\phi$-theory.

Let us rephrase the above problem in terms of polynomials. Let $q(X,Y)$ be the Laurent polynomial associated with $D$. By an appropriate translation, we can write $X^m Y^nq(X,Y) =  \tilde q(X,Y)$, where $\tilde q(X,Y)$ is a polynomial. If there is a polynomial $\tilde r(X,Y)$ such that $\tilde q(X,Y) = \tilde r(X,Y) \tilde p(X,Y)$, then $D$ is of higher order than $\Delta_L$, i.e., $D=D'\circ\Delta_L$, where $D'$ is the operator associated with $X^a Y^b\tilde r(X,Y)$ for some $a,b\in\mathbb Z$.

With the above preparations, the central result of this appendix can be stated in terms of polynomials as follows: suppose $\tilde q(X,Y)$ is a polynomial such that
\ie\label{polyphi-diffeq}
\tilde q(X,Y) \hat f(X,Y) = 0~,\quad\forall f\in\mathcal H(\mathbb Z^2,\mathbb R)~,
\fe
then $\tilde q(X,Y) = \tilde r(X,Y) \tilde p(X,Y)$ for some $\tilde r (X,Y)\in \mathbb{Z}[X,Y]$, where $\tilde p(X,Y)$ is the polynomial \eqref{polyptilde} associated with the discrete Laplacian operator $\Delta_L$.

More specifically, since $\tilde p(X,Y)$ is monic with leading term $X^2Y$, by multivariate division with respect to lexicographic order, $\tilde q(X,Y)$ can be written uniquely as
\ie\label{polyphi-multivardiv}
\tilde q(X,Y) = X^2 \alpha(X) + X \beta(Y) + \gamma(Y) + \tilde r(X,Y) \tilde p(X,Y)~,
\fe
where $\alpha(X)\in\mathbb Z[X]$, $\beta(Y),\gamma(Y)\in\mathbb Z[Y]$, and $\tilde r(X,Y)\in\mathbb Z[X,Y]$. The above statement is then equivalent to showing that $\alpha(X) = \beta(Y) = \gamma(Y) = 0$ if \eqref{polyphi-diffeq} is obeyed, which means that $D$ is of higher order than $\Delta_L$.

We parameterize the polynomials as
\ie
\alpha(X) = \sum_{i=0}^u a_i X^i~,\qquad \beta(Y)=\sum_{j=0}^v b_j Y^j~,\qquad \gamma(Y)=\sum_{k=0}^w c_k Y^k~,
\fe
with nonnegative integers $u$, $v$, and $w$.  Then, we apply \eqref{polyphi-multivardiv} to a specific set of discrete harmonic functions parameterized by $t$:\footnote{It is related to the \emph{discrete exponential function} on the square lattice $\mathbb Z^2$ \cite{Duffin1968TheDA,lovasz2004discrete}, which is the discrete analog the exponential function on the complex plane $\mathbb C \cong \mathbb R^2$.\label{ftnt:discexp}}
\ie\label{discexp}
\begin{split}
&f_t(x,y) \equiv X_t^xY_t^y~,
\\
&X_t \equiv -t\left(\frac{1+t}{1-t}\right)~,\qquad Y_t\equiv t\left(\frac{1-t}{1+t}\right)~.
\end{split}
\fe
(It is easy to check that $f_t\in\mathcal H(\mathbb Z^2,\mathbb R)$ for any $t\ne 0,\pm 1$, i.e., $\tilde p(X,Y) \hat f_t(X,Y) = 0$.)  Then, using \eqref{polyphi-multivardiv} and \eqref{polyphi-diffeq}, we get
\ie\label{polyphi-polyt1}
X_t^2 \alpha(X_t) + X_t \beta(Y_t) + \gamma(Y_t) = 0~.
\fe
Next, we turn this into a polynomial in $t$ by multiplying it by $(1-t)^{u+2} (1+t)^\nu$ with $\nu = \max(v-1,w)$
\ie\label{polyphi-polyt}
&\sum_{i=0}^u a_i \alpha_i(t;u,\nu) + \sum_{j=0}^v b_j \beta_j(t;u,\nu) + \sum_{k=0}^w c_k \gamma_k(t;u,\nu) = 0~,
\fe
where
\ie
&\alpha_i(t;u,\nu) = (1-t)^{u+2} (1+t)^\nu X_t^{i+2}= (-t)^{i+2} (1+t)^{\nu+i+2} (1-t)^{u-i}~,
\\
&\beta_j(t;u,\nu) = (1-t)^{u+2} (1+t)^\nu X_t Y_t^j = -t^{j+1} (1+t)^{\nu-j+1} (1-t)^{u+j+1}~,
\\
&\gamma_k(t;u,\nu) = (1-t)^{u+2} (1+t)^\nu Y_t^k = t^k (1+t)^{\nu-k} (1-t)^{u+k+2}~.
\fe
Since the equation \eqref{polyphi-polyt} holds for any $t\ne0,\pm 1$, the polynomial in \eqref{polyphi-polyt} must vanish identically, even at $t=0,\pm1$. What can we say about the coefficients $a_i$'s, $b_j$'s, and $c_k$'s then?

For fixed $(u,v,w)$, we have a set  $\mathcal P(u,v,w)$ of polynomials in $t$, $\{\alpha_i:i=0,\ldots,u\}\cup \{\beta_j:j=0,\ldots,v\}\cup\{\gamma_k:k=0,\ldots,w\}$.  We will show that these polynomials are linearly independent, and therefore, $\alpha(X) = \beta(Y) = \gamma(Y) = 0$ and $\tilde q(X,Y) = \tilde r(X,Y) \tilde p(X,Y)$.

First, note that for $v\le v_0$ and $w\le v_0-1$, we have $\mathcal P(u,v_0,w)\subseteq \mathcal P(u,v_0,v_0-1)$ and $\mathcal P(u,v,v_0-1) \subseteq \mathcal P(u,v_0,v_0-1)$ because $\max(v-1,v_0-1) = \max(v_0-1,w)$. So it suffices to show that the polynomials in the set $\mathcal P(u,v_0,v_0-1)$ are linearly independent for all $u\ge0$ and $v_0\ge1$. We proceed by induction:

\begin{itemize}
\item \underline{Base case}: It is easy to check that the set $\mathcal P(0,1,0)$ is linearly independent, and hence $\mathcal P(0,0,0)$ is also linearly independent.

\item \underline{Induction step}: Assume that $\mathcal P(u,v_0,v_0-1)$ is linearly independent. Consider $\mathcal P(u+1,v_0,v_0-1)$:
\ie
&\alpha_i(t;u+1,v_0-1) = \begin{cases}
(1-t)\alpha_i(t;u,v_0-1)~,&\text{for } i = 0,\ldots,u~,
\\
(-t)^{u+3} (1+t)^{v_0+u+2}~,&\text{for } i=u+1~,
\end{cases}
\\
&\beta_j(t;u+1,v_0-1) = (1-t) \beta_j(t;u,v_0-1)~,\quad\text{for } j = 0,\ldots,v_0~,
\\
&\gamma_k(t;u+1,v_0-1) = (1-t) \gamma_k(t;u,v_0-1)~,\quad\text{for } k = 0,\ldots,v_0-1~.
\fe
The polynomials in the first, third, and fourth lines are linearly independent by the induction hypothesis. The second line is nonzero at $t=1$, whereas the other three lines vanish at $t=1$, so the second line is independent of the other polynomials. Thus, $\mathcal P(u+1,v_0,v_0-1)$ is linearly independent.

Now consider $\mathcal P(u,v_0+1,v_0)$:
\ie
&\alpha_i(t;u,v_0) = (1+t)\alpha_i(t;u,v_0-1)~,\quad \text{for } i = 0,\ldots,u~,
\\
&\beta_j(t;u,v_0) = \begin{cases}
(1+t) \beta_j(t;u,v_0-1)~,&\text{for } j = 0,\ldots,v_0~,
\\
-t^{v_0+2}(1-t)^{u+v_0+2}~,&\text{for } j = v_0+1,
\end{cases}
\\
&\gamma_k(t;u,v_0) = \begin{cases}
(1+t) \gamma_k(t;u,v_0-1)~,&\text{for } k = 0,\ldots,v_0-1~,
\\
t^{v_0}(1-t)^{u+v_0+2}~,&\text{for } k = v_0,
\end{cases}
\fe
The polynomials in the first, second, and fourth lines are linearly independent by the induction hypothesis. Those in the third and fifth lines are linear independent of the other polynomials because they do not vanish at $t=-1$, and of each other because they have different degrees. Thus, $\mathcal P(u,v_0+1,v_0)$ is linearly independent.
\end{itemize}
Therefore, $\mathcal P(u,v,w)$ is linearly independent for any $(u,v,w)$. Since the polynomial in $t$ in \eqref{polyphi-polyt} must vanish identically, it follows that $a_i = b_j = c_k = 0$, so $\alpha(X)=\beta(Y)=\gamma(Y)=0$. Hence, $\tilde q(X,Y) = \tilde r(X,Y) \tilde p(X,Y)$, which is what we set out to show.

It follows that any difference operator $D$ (which is associated with the polynomial $\tilde q(X,Y)$) respecting the momentum global symmetry must be of higher order than $\Delta_L$, i.e., $D= D'\circ \Delta_L$.

Next, we impose also the winding symmetry.  This excludes terms such as $\cos\tilde \phi$.  Using an argument similar to the one above, it is easy to see that the model is also natural with respect to the winding symmetry.  One way to see that is to first dualize the theory and apply the argument above with $\phi \leftrightarrow \tilde \phi$.  We conclude that the action \eqref{2dlapphi-modVill-action} is natural if we impose its entire global symmetry.

\subsection{Mobility of defects in 2+1d $U(1)$ Laplacian gauge theory}\label{app:polyA}

In this appendix, we prove the immobility of any finite set of defects with arbitrary charges (except in some trivial cases) in the 2+1d $U(1)$ Laplacian gauge theory on the infinite square lattice $\mathbb Z^2$.

Before proceeding, let us distinguish between two kinds of defects that capture the motion of a particle. Typically, when a particle can move between two points, there is an operator supported in a small region, e.g., a line joining the two points. However, there are also situations where the operator that moves the particle can have a support spanning $O(L_x)$ or $O(L_y)$ sites even though the two points are separated by a much smaller distance. (See \cite{Gorantla:2022eem} for a discussion and examples of both kinds of operators.) The existence of the latter kind of operators depends on the number-theoretic properties of $L_i$, whereas the former kind of operators exist for all $L_i$. In particular, only the former make sense on the infinite square lattice.

Consider the defect
\ie\label{polyA-frac}
\exp\left[i \sum_{\tau}\mathcal A_\tau(\tau+\tfrac12,x,y) \right]~,
\fe
which describes the worldline of a single particle with unit charge. Under the time-like symmetry that shifts
\ie\label{polyA-lin}
\mathcal A_\tau(\tau+\tfrac12,x,y) \rightarrow \mathcal A_\tau(\tau+\tfrac12,x,y) + \delta_{\tau,0} \left( \frac{2\pi m_x x}{L_x} + \frac{2\pi m_y y}{L_y} \right)~,\qquad m_x, m_y \in \mathbb Z~,
\fe
the defect \eqref{polyA-frac} acquires an $(x,y)$-dependent phase, so it cannot bend. In other words, the particle is completely immobile, i.e., it is a fracton.

Since the time-like symmetry in \eqref{polyA-lin} is present also in the scalar charge theory, the same conclusion holds there. The selection rules from the time-like global symmetries give a more precise explanation of the intuitive “dipole moment conservation” discussed in \cite{Pretko:2016kxt,Pretko:2016lgv,Pretko:2018jbi}. (See \cite{Gorantla:2022eem} for a discussion.)

Next, consider the defect
\ie\label{polyA-dip}
\exp\left(i \sum_{\tau} \left[\mathcal A_\tau(\tau+\tfrac12,x+x_0,y+y_0) - \mathcal A_\tau(\tau+\tfrac12,x,y)\right] \right)~,
\fe
which describes the worldline of a dipole of particles with charges $\pm 1$ with separation $(x_0,y_0)$. The shift \eqref{polyA-lin} imposes the constraint that the defect cannot move unless the separation is held fixed. This is the only constraint in the scalar charge theory, and as long as it is met, the dipole can move. There are additional constraints in the Laplacian gauge theory. Indeed, under the time-like symmetry that shifts (for simplicity, we set $L_x = L_y = L$)
\ie
\mathcal A_\tau(\tau+\tfrac12,x,y) \rightarrow \mathcal A_\tau(\tau+\tfrac12,x,y) + \delta_{\tau,0} \left[\frac{2\pi m x y}{L} + \frac{2\pi m' (x^2 -y^2 - L x + L y)}{2L} \right]~,
\fe
where $m,m'\in\mathbb Z$, the defect \eqref{polyA-dip} acquires an $(x,y)$-dependent phase, so it cannot bend. In other words, the dipole is also completely immobile.

More generally, consider the defect
\ie\label{polyA-def}
\exp\left[ i \sum_{\tau} \sum_{i=1}^n q_i \mathcal A_\tau(\tau+\tfrac12,x_i,y_i) \right]~.
\fe
which describes the world-lines of $n$ particles labelled by $i=1,\ldots,n$, with positions $(x_i,y_i)$, and charges $q_i$. It is difficult to analyze this case in full generality on a finite lattice, so we limit ourselves to an infinite square lattice.

Under the shift of $\mathcal A_\tau$ by a discrete harmonic function $f\in\mathcal H(\mathbb Z^2,\mathbb R)$ at a fixed time $\tau = 0$, the phase acquired by the defect \eqref{polyA-def} is $\exp\left[i\sum_{i=1}^n q_i f(x_i,y_i)\right]$. The defect carries trivial time-like charges (i.e., it is in the trivial superselection sector) if and only if for all discrete harmonic functions $f\in\mathcal H(\mathbb Z^2,\mathbb R)$
\ie\label{polyA-triv}
\sum_{i=1}^n q_i f(x_i,y_i) = 0\quad \iff \quad q(X,Y) \hat f(X,Y) = 0~.
\fe
(Once again, we cannot impose the weaker condition $\sum_{i=1}^n q_i f(x_i,y_i) = 0 \mod 2\pi$ because this equation should be true even for $c f(x,y)$ for any $c\in\mathbb R$.) As we showed in Section \ref{app:polyphi}, this is possible if and only if $q(X,Y) = r(X,Y) p(X,Y)$ for some Laurent polynomial $r(X,Y)$.

To see the physical meaning of being invariant under the time-like symmetry, assume that such an $r(X,Y)$ exists. Then, we can construct the following defect
\ie\label{polyA-defanni}
\exp\left[ i \sum_{\tau<0} \sum_{i=1}^n q_i \mathcal A_\tau(\tau+\tfrac12,x_i,y_i) \right] \times \exp\left[ -i\sum_{j=1}^m r_j \mathcal A(0,x_j,y_j) \right]~.
\fe
Here, $r_j$ and $(x_j,y_j)$ are obtained from $r(X,Y) = \sum_{j=1}^m r_j X^{x_j} Y^{y_j}$. This defect describes annihilation of the $n$ particles at time $\tau = 0$. To see that this defect is gauge invariant, observe that, under a gauge transformation, the exponent transforms as
\ie
\sum_{i=1}^n q_i \alpha(0,x_i,y_i) - \sum_{j=1}^m r_j \Delta_L \alpha(0,x_j,y_j)~.
\fe
This is the coefficient of $X^0 Y^0$ term in $[q(X,Y) - r(X,Y) p(X,Y)] \hat \alpha(X,Y)$, and so it vanishes.\footnote{Here, $\hat \alpha(X,Y)$ is the formal Laurent power series associated with the gauge parameter $\alpha(x,y)$. It should not be confused with the gauge parameter $\hat \alpha_i$ in Section \ref{sec:2ddipphi}.}

In fact, we can write the defect \eqref{polyA-defanni} as
\ie\label{polyA-localanni}
&\exp\left[ i \sum_{\tau<0} \sum_{j=1}^m r_j \Delta_L\mathcal A_\tau(\tau+\tfrac12,x_j,y_j) \right] \times \exp\left[ -i\sum_{j=1}^m r_j \mathcal A(0,x_j,y_j) \right]
\\
&=\prod_{j=1}^m \exp\left[ i r_j \sum_{\tau<0} \Delta_L\mathcal A_\tau(\tau+\tfrac12,x_j,y_j) - i r_j\mathcal A(0,x_j,y_j) \right]~.
\fe
Each factor here describes particles being annihilated ``locally'' because the operator that annihilates them is local.

The result in \eqref{polyA-localanni} can be understood intuitively as follows.  In this special case, the collection of defects coming from the past \eqref{polyA-def} can be expressed as a ``total spatial derivative'' using the Laplacian as in the first factor in \eqref{polyA-localanni}.  In this form, each term with the Laplacian can end using the local operator made out of $\mathcal A$ in the second factor in \eqref{polyA-localanni}.  The result, in this case, is that the collection of defects can be annihilated by an operator at time $\tau=0$.\footnote{This is analogous to the following very well known fact in standard $U(1)$ gauge theories. A dipole of particles with charges $\pm 1$ is represented by the defect $\exp(i \int d\tau~[A_\tau(\tau, x) -A_\tau (\tau,0)]) = \exp[i \int d\tau \int_0^x dx' ~\partial_x A_\tau(\tau,x')]$.  It can end at $\tau=0$, as described by $\exp(i \int_{-\infty}^0 d\tau~[A_\tau(\tau, x) -A_\tau (\tau,0)]) \times \exp[-i \int_0^x dx'~A_x(0,x')]$.}

Let us now examine the mobility of the $n$ particles described by the defect \eqref{polyA-def}. We stress that we consider mobility only under the restriction that the charges of the particles and the separations between them remain fixed. (Relaxing these two restrictions can lead to more possibilities, which we will not discuss here.) Then, we say that the $n$ particles can move by $(x_0,y_0)\ne (0,0)$ if there is a defect of the form
\ie\label{polyA-defmove}
&\exp\left[ i \sum_{\tau<0} \sum_{i=1}^n q_i \mathcal A_\tau(\tau+\tfrac12,x_i,y_i) \right] \times \exp\left[ i\sum_{k=1}^{l} s_k \mathcal A(0,x_k,y_k) \right]
\\
&\qquad \times \exp\left[ i \sum_{\tau \ge 0} \sum_{i=1}^n q_i \mathcal A_\tau(\tau+\tfrac12,x_i+x_0,y_i+y_0) \right]~.
\fe
This defect exists, i.e., it is gauge invariant, if and only if
\ie\label{polyA-mob}
(X^{x_0} Y^{y_0} -1) q(X,Y) = s(X,Y) p(X,Y)~,
\fe
where $s(X,Y) = \sum_{k=1}^l s_k X^{x_k} Y^{y_k}$. Equivalently, this is precisely the condition for which the time-like charges of the $n$ particles remain unchanged after displacing them by $(x_0,y_0)$.

If $q(X,Y)$ is a multiple of $p(X,Y)$, i.e., $q(X,Y) = r(X,Y) p(X,Y)$ for some Laurent polynomial $r(X,Y)$, then we can always choose $s(X,Y) = r(X,Y) (X^{x_0} Y^{y_0} - 1)$ so that \eqref{polyA-mob} is satisfied. However, this is not an interesting situation because, when $q(X,Y)= r(X,Y) p(X,Y)$, the defect \eqref{polyA-def} has trivial time-like charges as explained around \eqref{polyA-triv}. Consequently, similar to the discussion around \eqref{polyA-localanni}, this situation can be interpreted as ``locally annihilating'' the particles and then ``locally creating'' them elsewhere. For example, when $r(X,Y) = 1$, the defect \eqref{polyA-defmove} is
\ie\label{polyA-trivdefmove}
&\exp\left[ i \sum_{\tau<0} \Delta_L \mathcal A_\tau(\tau+\tfrac12,0,0) \right] \times \exp\left[ i \mathcal A(0,x_0,y_0) - i\mathcal A(0,0,0) \right]
\\
&\qquad \times \exp\left[ i \sum_{\tau \ge 0} \Delta_L \mathcal A_\tau(\tau+\tfrac12,x_0,y_0) \right]~,
\fe
where the operator $e^{-i\mathcal A(0,0,0)}$ annihilates the particles around $(0,0)$ and then the operator $e^{i\mathcal A(0,x_0,y_0)}$ creates them around $(x_0,y_0)$. For more general $r(X,Y)$, the defect \eqref{polyA-defmove} is a product of defects of the form \eqref{polyA-trivdefmove}.

Can we have a defect like \eqref{polyA-defmove} when $q(X,Y)$ is not a multiple of $p(X,Y)$? Imposing \eqref{polyA-mob}, we see that this can happen if and only if $X^{x_0} Y^{y_0} - 1$ shares a nontrivial factor with $p(X,Y)$. Let us show that the latter cannot happen.

First, it is easy to check that $p(X,Y)$ is monic, non-constant, and irreducible up to a monomial.\footnote{A polynomial is said to be \emph{irreducible} if it cannot be written as product of two polynomials, neither of which is $\pm1$. We say a Laurent polynomial $g(X,Y)$ is \emph{irreducible up to a monomial} if $X^aY^bg(X,Y)$ is an irreducible polynomial for some $a,b\in\mathbb Z$. For example, $\tilde p(X,Y) = XY p(X,Y)$ is an irreducible polynomial, so $p(X,Y)$ is irreducible up to a monomial.} Let $d=\gcd(x_0,y_0)$, which is well defined because $(x_0,y_0)\ne (0,0)$. We can write
\ie\label{polyphi2:factor}
X^{x_0} Y^{y_0} -1 = (X^{x_0'} Y^{y_0'})^d -1 = (X^{x_0'} Y^{y_0'} -1) t(X,Y)~,
\fe
where $x_0'=x_0/d$, $y_0'=y_0/d$, and $t(X,Y) = \sum_{c=0}^{d-1} (X^{x_0'} Y^{y_0'})^c$ is a Laurent polynomial with $t(1,1)=d\ne 0$. The last condition implies that $p(X,Y)$ cannot share a nontrivial factor with $t(X,Y)$. Since $\gcd(x_0',y_0')=1$, the factor $X^{x_0'} Y^{y_0'} -1$ is monic, non-constant, and irreducible up to a monomial \cite{GAO2001501}. So, $p(X,Y)$ cannot share a nontrivial factor with $X^{x_0'} Y^{y_0'} -1$ as well. Therefore, $p(X,Y)$ does not share a nontrivial factor with $X^{x_0} Y^{y_0} - 1$.

To summarize, a finite set of charged particles cannot move in the 2+1d $U(1)$ Laplacian gauge theory on a square lattice unless they are in the trivial superselection sector, i.e., they can be ``annihilated locally.'' We remind the reader that when we say ``move'', the particles retain their charges and move in a rigid way.

\bibliographystyle{JHEP_modified}
\bibliography{laplacian,fracton}

\providecommand{\href}[2]{#2}\begingroup\raggedright\begin{thebibliography}{10}

\bibitem{Chamon:2004lew}
C.~Chamon, {\it Quantum glassiness in strongly correlated clean systems: An
  example of topological overprotection},  {\em Phys. Rev. Lett.} {\bf 94},
  040402 (2005).

\bibitem{Haah:2011drr}
J.~Haah, {\it {Local stabilizer codes in three dimensions without string
  logical operators}},  {\em Phys. Rev. A} {\bf 83}, 042330 (2011).

\bibitem{Vijay:2016phm}
S.~Vijay, J.~Haah, and L.~Fu, {\it {Fracton Topological Order, Generalized
  Lattice Gauge Theory and Duality}},  {\em Phys. Rev. B} {\bf 94}, 235157
  (2016).

\bibitem{Nandkishore:2018sel}
R.~M. Nandkishore and M.~Hermele, {\it {Fractons}},  {\em Ann. Rev. Condensed
  Matter Phys.} {\bf 10}, 295--313 (2019).

\bibitem{Pretko:2020cko}
M.~Pretko, X.~Chen, and Y.~You, {\it {Fracton Phases of Matter}},  {\em Int. J.
  Mod. Phys. A} {\bf 35}, 2030003 (2020).

\bibitem{Grosvenor:2021hkn}
K.~T. Grosvenor, C.~Hoyos, F.~Pe\~na Benitez, and P.~Sur\'owka, {\it
  {Space-Dependent Symmetries and Fractons}},  {\em Front. in Phys.} {\bf 9},
  792621 (2022).

\bibitem{Brauner:2022rvf}
T.~Brauner, S.~A. Hartnoll, P.~Kovtun, H.~Liu, M.~Mezei, A.~Nicolis, R.~Penco,
  S.-H. Shao, and D.~T. Son, {\it {Snowmass White Paper: Effective Field
  Theories for Condensed Matter Systems}},  in {\em {2022 Snowmass Summer
  Study}}, 3, 2022.
\newblock \href{http://arxiv.org/abs/2203.10110}{{\tt arXiv:2203.10110}}.

\bibitem{McGreevy:2022oyu}
J.~McGreevy, {\it {Generalized Symmetries in Condensed Matter}},
  \href{http://arxiv.org/abs/2204.03045}{{\tt arXiv:2204.03045}}.

\bibitem{Cordova:2022ruw}
C.~Cordova, T.~T. Dumitrescu, K.~Intriligator, and S.-H. Shao, {\it {Snowmass
  White Paper: Generalized Symmetries in Quantum Field Theory and Beyond}},  in
  {\em {2022 Snowmass Summer Study}}, 5, 2022.
\newblock \href{http://arxiv.org/abs/2205.09545}{{\tt arXiv:2205.09545}}.

\bibitem{Haah:2020ghp}
J.~Haah, {\it {A degeneracy bound for homogeneous topological order}},  {\em
  SciPost Phys.} {\bf 10}, 011 (2021).

\bibitem{Henley1997}
C.~L. Henley, {\it Relaxation time for a dimer covering with height
  representation},  {\em Journal of Statistical Physics} {\bf 89}, 483--507
  (Nov, 1997).

\bibitem{Moessner2001}
R.~Moessner, S.~L. Sondhi, and E.~Fradkin, {\it Short-ranged resonating valence
  bond physics, quantum dimer models, and ising gauge theories},  {\em Physical
  Review B} {\bf 65} (Dec, 2001).

\bibitem{Vishwanath:2004}
A.~Vishwanath, L.~Balents, and T.~Senthil, {\it Quantum criticality and
  deconfinement in phase transitions between valence bond solids},  {\em
  Physical Review B} {\bf 69} (Jun, 2004).

\bibitem{Fradkin:2004}
E.~Fradkin, D.~A. Huse, R.~Moessner, V.~Oganesyan, and S.~L. Sondhi, {\it
  Bipartite rokhsar--kivelson points and cantor deconfinement},  {\em Physical
  Review B} {\bf 69} (Jun, 2004).

\bibitem{Ardonne:2003wa}
E.~Ardonne, P.~Fendley, and E.~Fradkin, {\it {Topological order and conformal
  quantum critical points}},  {\em Annals Phys.} {\bf 310}, 493--551 (2004).

\bibitem{Ghaemi2005}
P.~Ghaemi, A.~Vishwanath, and T.~Senthil, {\it Finite-temperature properties of
  quantum lifshitz transitions between valence-bond solid phases: An example of
  local quantum criticality},  {\em Physical Review B} {\bf 72}, 024420 (Jul,
  2005).

\bibitem{Chen:2009ka}
B.~Chen and Q.-G. Huang, {\it {Field Theory at a Lifshitz Point}},  {\em Phys.
  Lett. B} {\bf 683}, 108--113 (2010).

\bibitem{2018PhRvB..98l5105M}
H.~{Ma} and M.~{Pretko}, {\it {Higher-rank deconfined quantum criticality at
  the Lifshitz transition and the exciton Bose condensate}},  {\em Physical
  Review B} {\bf 98}, 125105 (Sept., 2018).

\bibitem{Yuan:2019geh}
J.-K. Yuan, S.~A. Chen, and P.~Ye, {\it {Fractonic Superfluids}},  {\em Phys.
  Rev. Res.} {\bf 2}, 023267 (2020).

\bibitem{Lake:2022ico}
E.~Lake, M.~Hermele, and T.~Senthil, {\it {Dipolar Bose-Hubbard model}},  {\em
  Phys. Rev. B} {\bf 106}, 064511 (2022).

\bibitem{Gorantla:2022eem}
P.~Gorantla, H.~T. Lam, N.~Seiberg, and S.-H. Shao, {\it {Global dipole
  symmetry, compact Lifshitz theory, tensor gauge theory, and fractons}},  {\em
  Phys. Rev. B} {\bf 106}, 045112 (2022).

\bibitem{Griffin:2013dfa}
T.~Griffin, K.~T. Grosvenor, P.~Horava, and Z.~Yan, {\it {Multicritical
  Symmetry Breaking and Naturalness of Slow Nambu-Goldstone Bosons}},  {\em
  Phys. Rev. D} {\bf 88}, 101701 (2013).

\bibitem{Griffin:2014bta}
T.~Griffin, K.~T. Grosvenor, P.~Horava, and Z.~Yan, {\it {Scalar Field Theories
  with Polynomial Shift Symmetries}},  {\em Commun. Math. Phys.} {\bf 340},
  985--1048 (2015).

\bibitem{Pretko:2016kxt}
M.~Pretko, {\it {Subdimensional Particle Structure of Higher Rank U(1) Spin
  Liquids}},  {\em Phys. Rev. B} {\bf 95}, 115139 (2017).

\bibitem{Pretko:2016lgv}
M.~Pretko, {\it {Generalized Electromagnetism of Subdimensional Particles: A
  Spin Liquid Story}},  {\em Phys. Rev. B} {\bf 96}, 035119 (2017).

\bibitem{Pretko:2018jbi}
M.~Pretko, {\it {The Fracton Gauge Principle}},  {\em Phys. Rev. B} {\bf 98},
  115134 (2018).

\bibitem{Gromov:2018nbv}
A.~Gromov, {\it {Towards classification of Fracton phases: the multipole
  algebra}},  {\em Phys. Rev. X} {\bf 9}, 031035 (2019).

\bibitem{Seiberg:2019vrp}
N.~Seiberg, {\it {Field Theories With a Vector Global Symmetry}},  {\em SciPost
  Phys.} {\bf 8}, 050 (2020).

\bibitem{Shenoy:2019wng}
V.~B. Shenoy and R.~Moessner, {\it {$(k,n)$-fractonic Maxwell theory}},  {\em
  Phys. Rev. B} {\bf 101}, 085106 (2020).

\bibitem{Gromov:2020rtl}
A.~Gromov, {\it {A Duality Between $U(1)$ Haah Code and 3D Smectic A Phase}},
  \href{http://arxiv.org/abs/2002.11817}{{\tt arXiv:2002.11817}}.

\bibitem{Gromov:2020yoc}
A.~Gromov, A.~Lucas, and R.~M. Nandkishore, {\it {Fracton hydrodynamics}},
  {\em Phys. Rev. Res.} {\bf 2}, 033124 (2020).

\bibitem{Chen:2020jew}
S.~A. Chen, J.-K. Yuan, and P.~Ye, {\it {Fractonic superfluids. II. Condensing
  subdimensional particles}},  {\em Phys. Rev. Res.} {\bf 3}, 013226 (2021).

\bibitem{Du:2021pbc}
Y.-H. Du, U.~Mehta, D.~X. Nguyen, and D.~T. Son, {\it {Volume-preserving
  diffeomorphism as nonabelian higher-rank gauge symmetry}},  {\em SciPost
  Phys.} {\bf 12}, 050 (2022).

\bibitem{Stahl:2021sgi}
C.~Stahl, E.~Lake, and R.~Nandkishore, {\it {Spontaneous breaking of multipole
  symmetries}},  {\em Phys. Rev. B} {\bf 105}, 155107 (2022).

\bibitem{Jensen:2022iww}
K.~Jensen and A.~Raz, {\it {Large $N$ fractons}},
  \href{http://arxiv.org/abs/2205.01132}{{\tt arXiv:2205.01132}}.

\bibitem{Kapustin:2022fzp}
A.~Kapustin and L.~Spodyneiko, {\it {Hohenberg-Mermin-Wagner-type theorems and
  dipole symmetry}},  \href{http://arxiv.org/abs/2208.09056}{{\tt
  arXiv:2208.09056}}.

\bibitem{Gorantla:2022mrp}
P.~Gorantla, H.~T. Lam, and S.-H. Shao, {\it {Fractons on Graphs and
  Complexity}},  \href{http://arxiv.org/abs/2207.08585}{{\tt
  arXiv:2207.08585}}.

\bibitem{Villain:1974ir}
J.~Villain, {\it {Theory of one-dimensional and two-dimensional magnets with an
  easy magnetization plane. 2. The Planar, classical, two-dimensional magnet}},
   {\em J. Phys. (France)} {\bf 36}, 581--590 (1975).

\bibitem{paper1}
N.~Seiberg and S.-H. Shao, {\it {Exotic Symmetries, Duality, and Fractons in
  2+1-Dimensional Quantum Field Theory}},  {\em SciPost Phys.} {\bf 10}, 027
  (2021).

\bibitem{paper2}
N.~Seiberg and S.-H. Shao, {\it {Exotic $U(1)$ Symmetries, Duality, and
  Fractons in 3+1-Dimensional Quantum Field Theory}},  {\em SciPost Phys.} {\bf
  9}, 046 (2020).

\bibitem{paper3}
N.~Seiberg and S.-H. Shao, {\it {Exotic $\mathbb{Z}_N$ Symmetries, Duality, and
  Fractons in 3+1-Dimensional Quantum Field Theory}},  {\em SciPost Phys.} {\bf
  10}, 003 (2021).

\bibitem{Gorantla:2020xap}
P.~Gorantla, H.~T. Lam, N.~Seiberg, and S.-H. Shao, {\it {More Exotic Field
  Theories in 3+1 Dimensions}},  {\em SciPost Phys.} {\bf 9}, 073 (2020).

\bibitem{Gorantla:2020jpy}
P.~Gorantla, H.~T. Lam, N.~Seiberg, and S.-H. Shao, {\it {fcc lattice,
  checkerboards, fractons, and quantum field theory}},  {\em Phys. Rev. B} {\bf
  103}, 205116 (2021).

\bibitem{Rudelius:2020kta}
T.~Rudelius, N.~Seiberg, and S.-H. Shao, {\it {Fractons with Twisted Boundary
  Conditions and Their Symmetries}},  {\em Phys. Rev. B} {\bf 103}, 195113
  (2021).

\bibitem{Gorantla:2021svj}
P.~Gorantla, H.~T. Lam, N.~Seiberg, and S.-H. Shao, {\it {A modified Villain
  formulation of fractons and other exotic theories}},  {\em J. Math. Phys.}
  {\bf 62}, 102301 (2021).

\bibitem{Gorantla:2021bda}
P.~Gorantla, H.~T. Lam, N.~Seiberg, and S.-H. Shao, {\it {Low-energy limit of
  some exotic lattice theories and UV/IR mixing}},  {\em Phys. Rev. B} {\bf
  104}, 235116 (2021).

\bibitem{Burnell:2021reh}
F.~J. Burnell, T.~Devakul, P.~Gorantla, H.~T. Lam, and S.-H. Shao, {\it
  {Anomaly inflow for subsystem symmetries}},  {\em Phys. Rev. B} {\bf 106},
  085113 (2022).

\bibitem{Sulejmanpasic:2019ytl}
T.~Sulejmanpasic and C.~Gattringer, {\it {Abelian gauge theories on the
  lattice: $\theta$-Terms and compact gauge theory with(out) monopoles}},  {\em
  Nucl. Phys. B} {\bf 943}, 114616 (2019).

\bibitem{Xu:2006}
C.~Xu, {\it Gapless bosonic excitation without symmetry breaking: An algebraic
  spin liquid with soft gravitons},  {\em Physical Review B} {\bf 74} (Dec,
  2006).

\bibitem{Bulmash:2018lid}
D.~Bulmash and M.~Barkeshli, {\it {The Higgs Mechanism in Higher-Rank Symmetric
  $U(1)$ Gauge Theories}},  {\em Phys. Rev. B} {\bf 97}, 235112 (2018).

\bibitem{Oh:2021gee}
Y.-T. Oh, J.~Kim, E.-G. Moon, and J.~H. Han, {\it {Rank-2 toric code in two
  dimensions}},  {\em Phys. Rev. B} {\bf 105}, 045128 (2022).

\bibitem{Oh:2022klh}
Y.-T. Oh, J.~Kim, and J.~H. Han, {\it {Effective Field Theory of Dipolar
  Braiding Statistics in Two Dimensions}},
  \href{http://arxiv.org/abs/2204.01279}{{\tt arXiv:2204.01279}}.

\bibitem{Manoj:2020bcz}
N.~Manoj, R.~Moessner, and V.~B. Shenoy, {\it {Fractonic View of Folding and
  Tearing Paper: Elasticity of Plates is Dual to a Gauge Theory with Vector
  Charges}},  {\em Phys. Rev. Lett.} {\bf 127}, 067601 (2021).

\bibitem{Pretko:2018qru}
M.~Pretko and L.~Radzihovsky, {\it {Fracton-Elasticity Duality}},  {\em Phys.
  Rev. Lett.} {\bf 120}, 195301 (2018).

\bibitem{Slagle:2018kqf}
K.~Slagle, A.~Prem, and M.~Pretko, {\it {Symmetric Tensor Gauge Theories on
  Curved Spaces}},  {\em Annals Phys.} {\bf 410}, 167910 (2019).

\bibitem{Pretko:2019omh}
M.~Pretko, Z.~Zhai, and L.~Radzihovsky, {\it {Crystal-to-Fracton Tensor Gauge
  Theory Dualities}},  {\em Phys. Rev. B} {\bf 100}, 134113 (2019).

\bibitem{Nguyen:2020yve}
D.~X. Nguyen, A.~Gromov, and S.~Moroz, {\it {Fracton-elasticity duality of
  two-dimensional superfluid vortex crystals: defect interactions and quantum
  melting}},  {\em SciPost Phys.} {\bf 9}, 076 (2020).

\bibitem{Gorantla:2022pii}
P.~Gorantla, H.~T. Lam, N.~Seiberg, and S.-H. Shao, {\it {Gapped lineon and
  fracton models on graphs}},  {\em Phys. Rev. B} {\bf 107}, 125121 (2023).

\bibitem{Ma:2018nhd}
H.~Ma, M.~Hermele, and X.~Chen, {\it {Fracton topological order from the Higgs
  and partial-confinement mechanisms of rank-two gauge theory}},  {\em Phys.
  Rev. B} {\bf 98}, 035111 (2018).

\bibitem{Pace2022}
S.~D. Pace and X.-G. Wen, {\it Position-dependent excitations and uv/ir mixing
  in the $\mathbb{Z}_{N}$ rank-2 toric code and its low-energy effective field
  theory},  \href{http://arxiv.org/abs/2204.07111}{{\tt arXiv:2204.07111}}.

\bibitem{Ebisu:2022nln}
H.~Ebisu and B.~Han, {\it {Anisotropic higher rank $\mathbb{Z}_N$ topological
  phases on graphs}},  \href{http://arxiv.org/abs/2209.07987}{{\tt
  arXiv:2209.07987}}.

\bibitem{Gaiotto:2014kfa}
D.~Gaiotto, A.~Kapustin, N.~Seiberg, and B.~Willett, {\it {Generalized Global
  Symmetries}},  {\em JHEP} {\bf 02}, 172 (2015).

\bibitem{BSMF:97}
R.~Bacher, P.~d. La~Harpe, and T.~Nagnibeda, {\it The lattice of integral flows
  and the lattice of integral cuts on a finite graph},  {\em Bulletin de la
  Soci\'et\'e Math\'ematique de France} {\bf 125}, 167--198 (1997).

\bibitem{BAKER2007}
M.~Baker and S.~Norine, {\it Riemann--roch and abel--jacobi theory on a finite
  graph},  {\em Advances in Mathematics} {\bf 215}, 766--788 (2007).

\bibitem{rasmussen}
A.~{Rasmussen}, Y.-Z. {You}, and C.~{Xu}, {\it {Stable Gapless Bose Liquid
  Phases without any Symmetry}},  {\em arXiv e-prints} arXiv:1601.08235 (Jan.,
  2016).

\bibitem{Pretko:2017xar}
M.~Pretko, {\it {Higher-Spin Witten Effect and Two-Dimensional Fracton
  Phases}},  {\em Phys. Rev. B} {\bf 96}, 125151 (2017).

\bibitem{POLYAKOV1977429}
A.~Polyakov, {\it Quark confinement and topology of gauge theories},  {\em
  Nuclear Physics B} {\bf 120}, 429 -- 458 (1977).

\bibitem{Yoneda:2022qpj}
M.~Yoneda, {\it {Equivalence of the modified Villain formulation and the dual
  Hamiltonian method in the duality of the XY-plaquette model}},
  \href{http://arxiv.org/abs/2211.01632}{{\tt arXiv:2211.01632}}.

\bibitem{Cheng:2022sgb}
M.~Cheng and N.~Seiberg, {\it {Lieb-Schultz-Mattis, Luttinger, and 't Hooft --
  anomaly matching in lattice systems}},
  \href{http://arxiv.org/abs/2211.12543}{{\tt arXiv:2211.12543}}.

\bibitem{Fazza:2022fss}
L.~Fazza and T.~Sulejmanpasic, {\it {Lattice quantum Villain Hamiltonians:
  compact scalars, U(1) gauge theories, fracton models and quantum Ising model
  dualities}},  {\em JHEP} {\bf 05}, 017 (2023).

\bibitem{Haah_2013}
J.~Haah, {\it Commuting pauli hamiltonians as maps between free modules},  {\em
  Communications in Mathematical Physics} {\bf 324}, 351--399 (oct, 2013).

\bibitem{tHooft:1979rat}
G.~'t~Hooft, {\it {Naturalness, chiral symmetry, and spontaneous chiral
  symmetry breaking}},  {\em NATO Sci. Ser. B} {\bf 59}, 135--157 (1980).

\bibitem{Duffin1968TheDA}
R.~J. Duffin and E.~L. Peterson, {\it The discrete analogue of a class of
  entire functions},  {\em Journal of Mathematical Analysis and Applications}
  {\bf 21}, 619--642 (1968).

\bibitem{lovasz2004discrete}
L.~Lov{\'a}sz, {\it Discrete analytic functions: an exposition},  {\em Surveys
  in Differential Geometry} {\bf 9}, 241--273 (2004).

\bibitem{GAO2001501}
S.~Gao, {\it Absolute irreducibility of polynomials via newton polytopes},
  {\em Journal of Algebra} {\bf 237}, 501--520 (2001).

\end{thebibliography}\endgroup

\end{document}